\documentclass[aps,amsfonts,pra,twocolumn,showpacs]{revtex4-1}

\usepackage{subfigure}
\usepackage{textcomp}
\usepackage{graphicx}
\usepackage{amssymb}
\usepackage{amsmath}
\usepackage{bm}
\usepackage{amsmath, amsthm, amssymb}
\usepackage{dsfont}
\usepackage[colorlinks=true,linkcolor=blue,urlcolor=blue,citecolor=blue]{hyperref}
\usepackage{multirow}
\usepackage{url}

\def\beq{\begin{equation}}
\def\eeq{\end{equation}}
\def\bea{\begin{eqnarray}}
\def\eea{\end{eqnarray}}

\begin{document}

\title{Operator growth and eigenstate entanglement in an interacting integrable Floquet system}

\author{Sarang Gopalakrishnan}
\affiliation{Department of Physics and Astronomy, CUNY College of Staten Island, Staten Island, NY 10314}
\affiliation{Physics Program and Initiative for the Theoretical Sciences, The Graduate Center, CUNY, New York, NY 10016}

\begin{abstract}

We analyze a simple model of quantum dynamics, which is a discrete-time deterministic version of the Frederickson-Andersen model. 
We argue that this model is integrable, with a quasiparticle description related to the classical hard-rod gas.
Despite the integrability of the model, commutators of physical operators grow as in generic chaotic models, with a diffusively broadening front, and local operators obey the eigenstate thermalization hypothesis (ETH). However, large subsystems violate ETH; as a function of subsystem size, eigenstate entanglement first increases linearly and then saturates at a scale that is parametrically smaller than half the system size. 

\end{abstract}

\maketitle

%\cite{cazalilla2010focus, polk_RMP, nhreview}. 
%

A central theme in many-body physics is the mechanics of thermalization, decoherence, and information scrambling in isolated, interacting quantum systems~\cite{cazalilla2010focus, polk_RMP, BAA, nhreview, langen2013local, kaufman2016quantum}. 
Generic systems (except for the many-body localized phase~\cite{BAA, nhreview}) are believed to obey the eigenstate thermalization hypothesis (ETH), which posits that local observables in a many-body eigenstate behave as they would in an appropriately chosen thermal state~\cite{deutsch_eth, srednicki_eth, Rigol:2008kq, ETH_outliers, dymarsky2017}. 
However, in one dimension, a number of physically important systems (e.g., the Heisenberg spin chain) are integrable---i.e., they have extensively many conservation laws and thus do not thermalize~\cite{Rigol:2008kq, ilievski2016}. The absence of thermalization in integrable systems, and the anomalously slow ``prethermal'' behavior of nearly integrable systems, have been experimentally observed~\cite{kinoshita, gring, langen2013local, lev_qnc, erne2018observation, li2018dephasing}. Some integrable systems can be solved by mapping to free fermions~\cite{sml_ising}, but many others, including the Heisenberg chain, cannot~\cite{faddeev}. In the latter class of ``interacting'' (i.e., Bethe-ansatz-solvable) integrable systems, it is challenging to compute the dynamics of physical observables, because physical operators have complicated representations in terms of the quasiparticles~\cite{caux_calab}, although coarse-grained approaches to some dynamical questions have recently been developed~\cite{ghd, fagotti2016, bethe_boltzmann, soliton_gases}. In the quantum context, most work on integrable systems has focused on Hamiltonian dynamics, though there has been some recent work on time-periodic, driven integrable systems~\cite{gritsev2017, vanicat2017}. 

This work presents and analyzes a simple integrable Floquet model for which many of these questions can be explicitly addressed. This model is a deterministic discrete-time version of the Frederickson-Andersen model~\cite{ritort2003, garrahan2017aspects, hickey2016}, which is a standard model of kinetically constrained dynamics; we call it the Floquet-Frederickson-Andersen (FFA) model. The dynamics of the FFA model is in some ways analogous to the classical hard-rod gas~\cite{spohn2012large, soliton_gases}, a canonical interacting integrable system; as we discuss, there is a natural description in terms of ballistically propagating quasiparticles.
Beyond its integrability, what renders the model tractable is that its dynamics maps each computational-basis product state to a unique computational-basis product state; this allows for efficient classical simulations of dynamics. Despite its simplicity the FFA model retains two key features of generic integrable systems: first, the relation between physical observables and quasiparticles is nontrivial, so physical observables are complicated in the quasiparticle language; and second, each quasiparticle's motion (and even the geometry it feels~\cite{spohn17}) is modified by the distribution of other quasiparticles. 

%The many-body eigenstates of the model are highly entangled, and operator entanglement also grows with time (in contrast to Clifford-gate models~\cite{gutschow2010, clclifford, nrvh, curtvonK}). The presence of a ``classical'' basis in our model allows us to efficiently construct eigenstates and compute their entanglement~\cite{gz}, and to explore the behavior of out-of-time-order commutators (OTOCs), for relatively large systems. 

\begin{figure}[b]
\begin{center}
\begin{minipage}{0.23\textwidth}
\includegraphics[width = \linewidth]{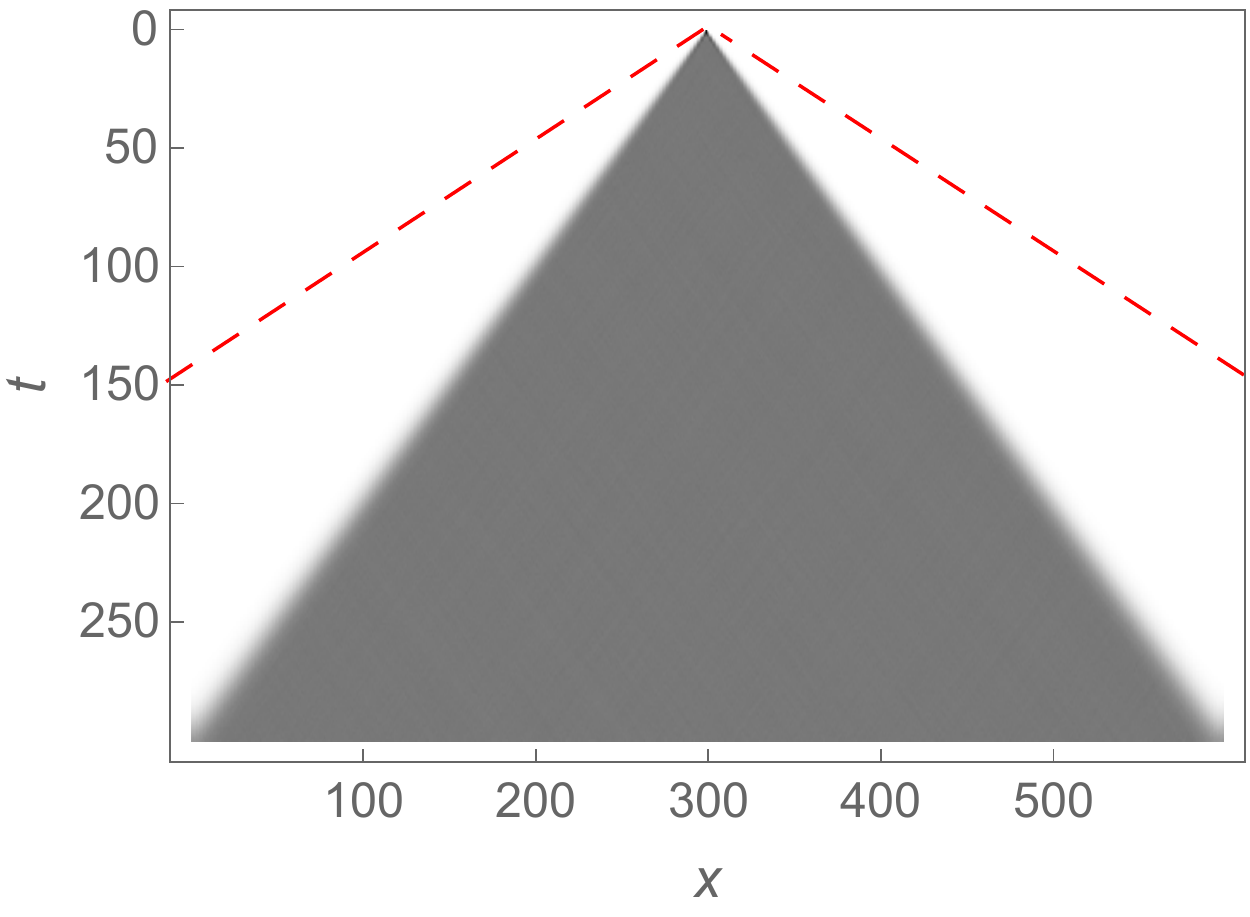}
\end{minipage}
\begin{minipage}{0.22\textwidth}
\includegraphics[width =  \linewidth]{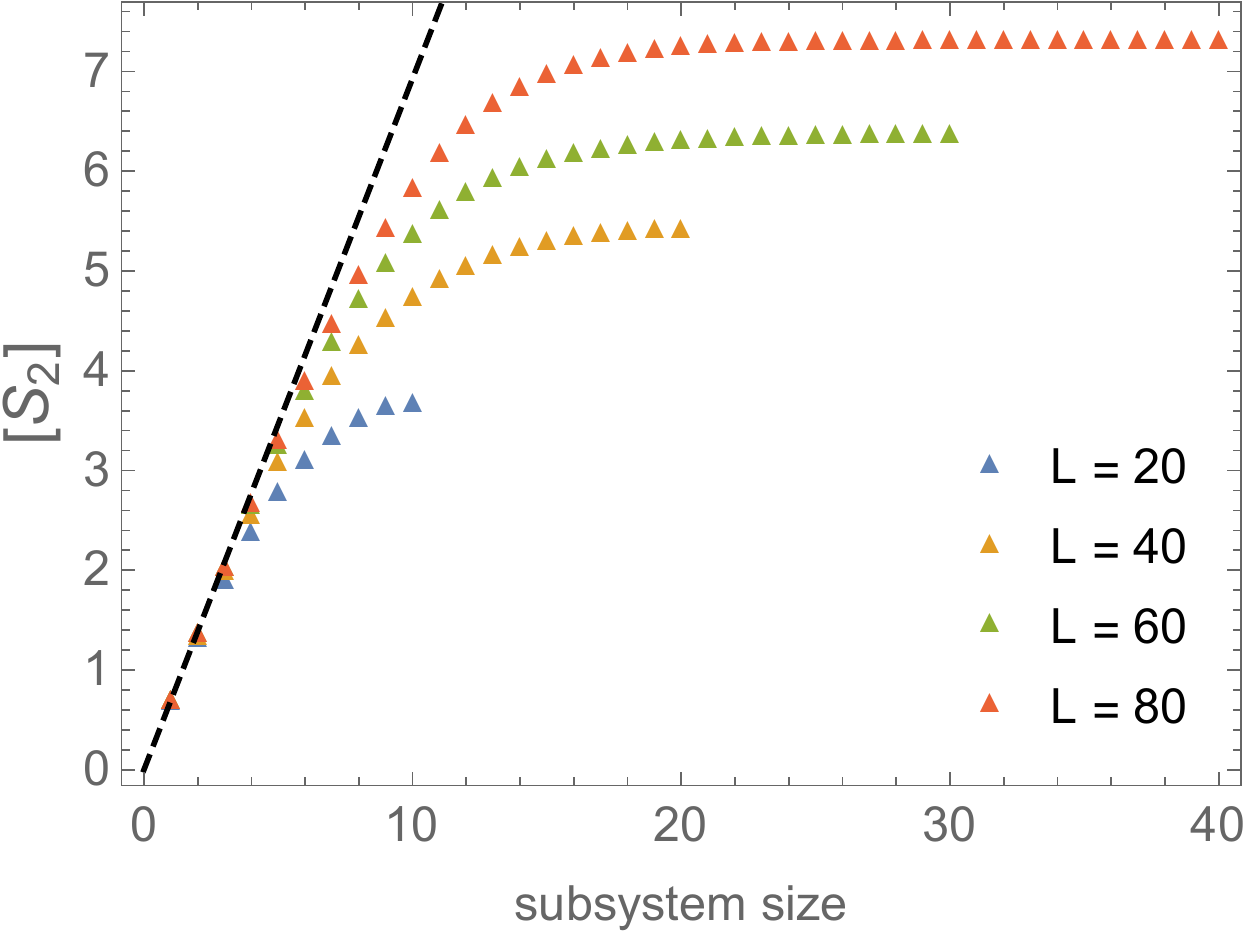}
\end{minipage}
\caption{Left: temporal growth of state-averaged out-of-time-order commutator for system size $L = 600$, averaged over $10000$ initial states. Dashed red lines show the causal light-cone velocity, outside which the commutator strictly vanishes. Right: second Renyi entropy $S_2$ vs. subsystem size, averaged over 30 random eigenstates, for various system sizes $L$. There is a clear crossover scale beyond which subsystem entanglement saturates.}
\label{fig1}
\end{center}
\end{figure}

Our main results are as follows (Fig.~\ref{fig1}). For physical observables the OTOC displays scrambling, with a front that broadens diffusively as expected for generic chaotic systems~\cite{xu2018a, xu2018b, khn}; however, this front is anomalous on timescales exceeding the system size $L$. In addition, small subsystems satisfy ETH, with a fraction of outlier states that appears to vanish (slowly) in the thermodynamic limit. However, large subsystems are strikingly nonthermal: for subsystem size $\agt 2 \ln L$, the eigenstate entanglement crosses over from a thermal volume law to a constant. This crossover sharpens for larger systems. %Finally, the off-diagonal matrix elements of local operators are distributed non-thermally: a local operator has $o(2^{-L/2})$ non-vanishing matrix elements between a given eigenstate and others.

\emph{Model}.---The model we consider was recently introduced in Ref.~\cite{gz}; the system is a spin-$1/2$ chain, subject to repeated application of the unitary

\beq\label{unitary}
U = W(\text{odd} \rightarrow \text{even}) W(\text{even} \rightarrow \text{odd}),
\eeq
where $W(\text{even} \rightarrow \text{odd})$ consists of the following rule, applied to each odd spin $n$: apply the Pauli operator $\sigma^x_n$ unless the two neighboring even sites, $n-1$ and $n+1$, are both in the $|\downarrow\rangle$ state. This rule can be composed from the gate sequence $\text{Toffoli}(n-1, n+1 \rightarrow n)\text{CNOT}(n - 1 \rightarrow n) \text{CNOT}(n+1 \rightarrow n)$, in which controlled NOT and Toffoli gates are applied to the target site from its two neighbors. $W(\text{odd} \rightarrow \text{even})$ repeats this process with even and odd sites exchanged. The full-cycle unitary $U$ has a strict causal light cone that expands at \emph{two} sites per period (dashed red lines in Fig.~\ref{fig1}). In what follows we treat periodic boundary conditions; the effects of other boundary conditions are deferred to future work. We also refer to odd (even) sites as the $A$ ($B$) sublattices respectively.

Eq.~\eqref{unitary} can be regarded as a reversible block cellular automaton~\cite{schumacher2004, jpg2018}. 
A number of works have addressed cellular automata and quantum circuits involving Clifford gates~\cite{gutschow2010, werner2010fractal, clclifford, nrvh, gz}. Clifford gates induce simple operator dynamics, mapping Pauli strings to other Pauli strings, and therefore do not give rise to operator entanglement~\cite{hosur2016chaos, jhn}. A related but Clifford-only model, without the Toffoli gate, was analyzed in Ref.~\cite{gz} and shown to map to free particles. In contrast, the Toffoli gate in Eq.~\eqref{unitary} induces operator entanglement for local operators, even after a single step of time evolution: for example, the operator $\sigma^x_i$ evolves to $\tilde{\sigma}^x_i \equiv \frac{1}{16}(1 + \sigma^x_{i - 1} + \sigma^z_{i - 2} - \sigma^z_{i - 2} \sigma^x_{i - 1}) \sigma^x_i (1 +\sigma^x_{i + 1}  + \sigma^z_{i+ 2}  - \sigma^x_{i + 1} \sigma^z_{i + 2})$, which cannot be factored into on-site operators. Our methods here cannot be used to compute operator entanglement, however, so we defer a full discussion to future work.

\begin{figure}[tb]
\begin{center}
\begin{minipage}{0.23\textwidth}
\includegraphics[width =  \linewidth]{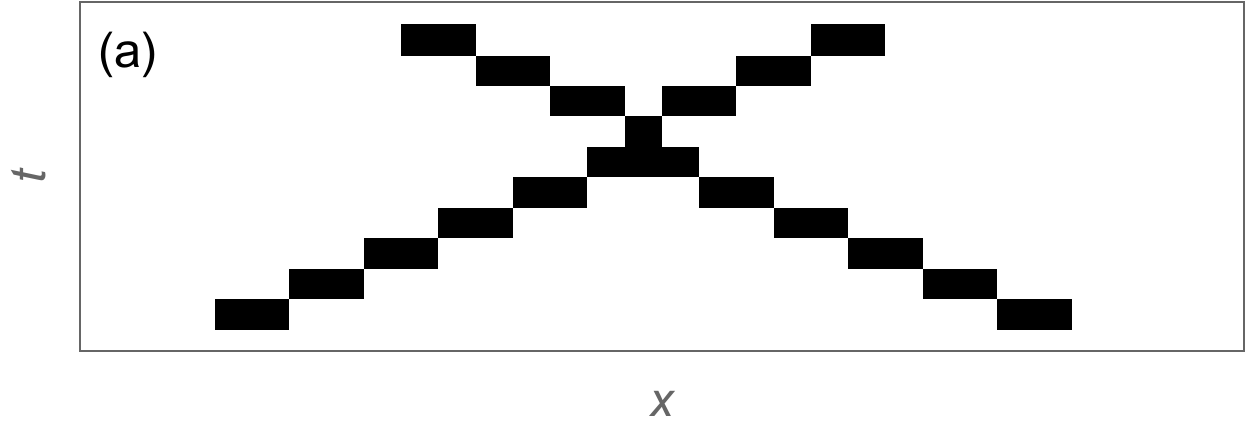}
\end{minipage}
\begin{minipage}{0.23\textwidth}
\includegraphics[width =   \linewidth]{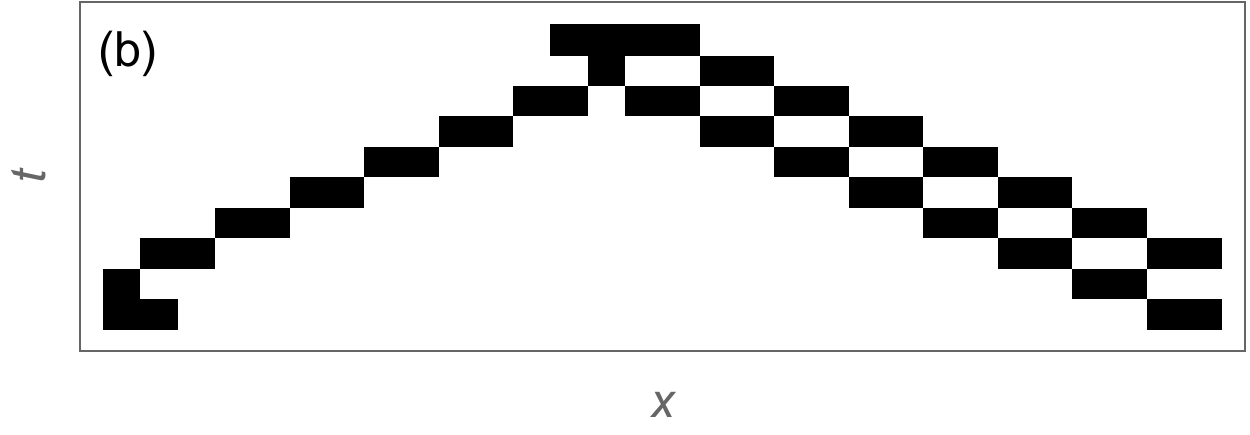}
\end{minipage}
\begin{minipage}{0.23\textwidth}
\includegraphics[width = \linewidth]{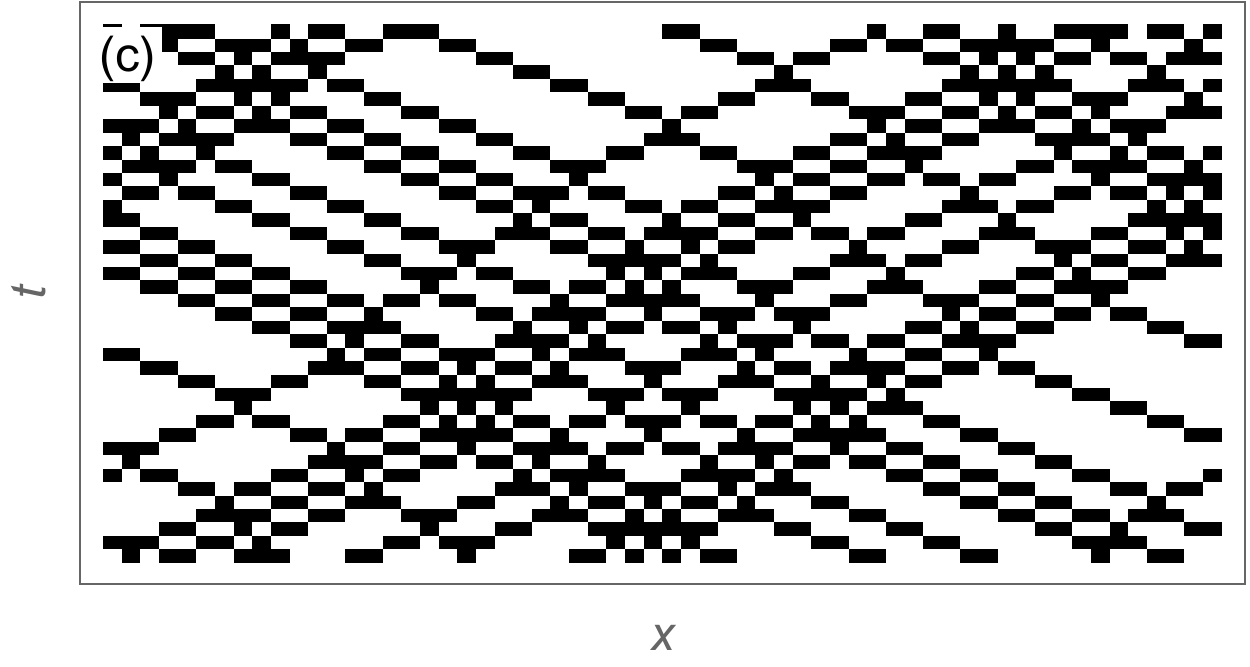}
\end{minipage}
\begin{minipage}{0.23\textwidth}
\includegraphics[width = \linewidth]{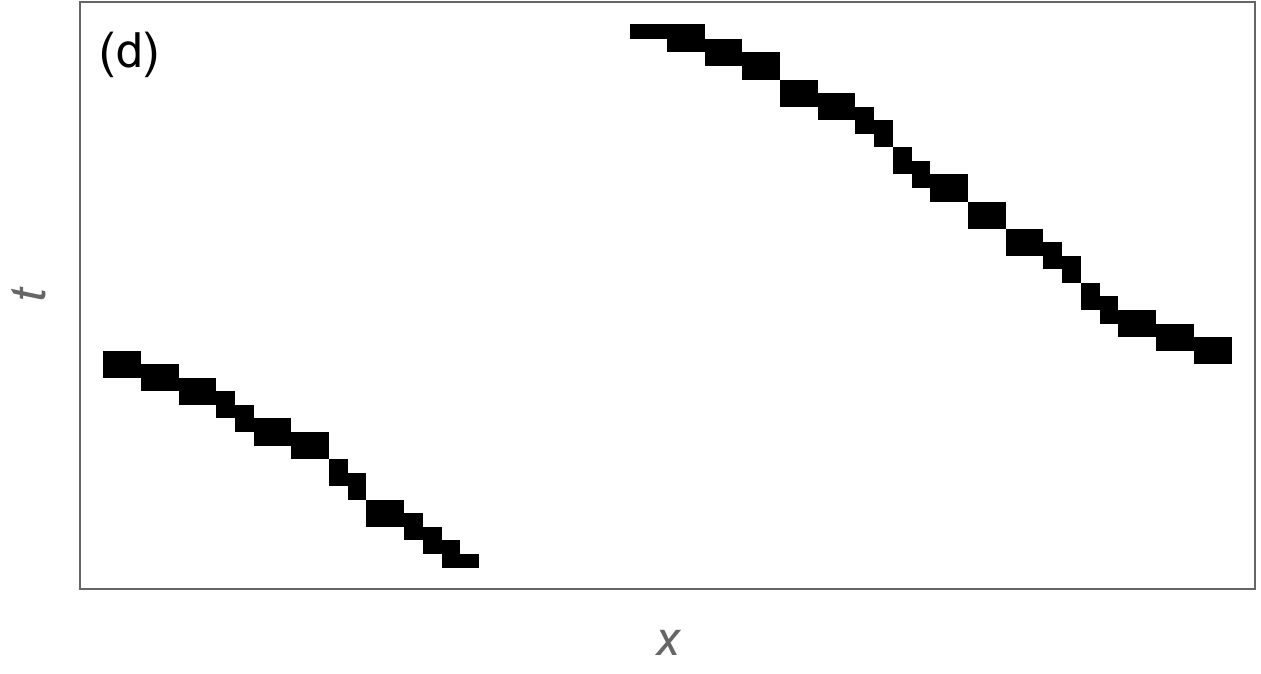}
\end{minipage}
\caption{Plots of spin dynamics; black cells are spin-up and white cells spin-down. (a)~Collision of a right-moving and left-moving quasiparticle. (b)~An initial state with four adjacent up spins is a three-quasiparticle state. (c)~Time evolution of a product state in which the left and right third are generic, whereas the middle third has only one quasiparticle. (d)~OTOC for this product state, obtained by moving one of the quasiparticles; note the absence of chaos.}
\label{qp}
\end{center}
\end{figure}

\emph{Quasiparticle picture}.---We first present a simple quasiparticle picture~\cite{yangyang} of the dynamics of this model; this picture is not derived microscopically, but is inferred from simulating the dynamics, and justified by its explanation of various exact numerical results.
%The dynamics of this model can be understood in terms of ballistically propagating quasiparticles. 
%
The structure of quasiparticles is easiest to describe when the density of up spins is low. In this limit, the elementary quasiparticles are pairs of two adjacent up spins surrounded by down spins. There are two inequivalent quasiparticles, respectively right- and left-moving, based on whether their sublattice structure is $AB$ or $BA$. (The dynamics is symmetric under exchanging sublattices and time-translating by half a period, but not under each action separately.) All right-movers and all left-movers have the same speed, so collisions necessarily involve a right-mover and a left-mover; also, each three-body collision can only occur in one sequence, precluding diffractive processes. 
Each collision induces a time delay of one time step, which is the same for all collisions [Fig.~\ref{qp}(a)]; thus, this model resembles hard rods with \emph{negative} rod length. The quantization condition for right (left) movers simply depends on the total number of left (right) movers.
Since all quasiparticles have the same velocity, a state is characterized by the \emph{spacings} between adjacent left-movers and between adjacent right-movers. Adjacent right-movers must have at least one empty site between them, since the configuration in Fig.~\ref{qp}(b) is not just a pair of right-movers; thus the state space for right-movers is constrained, with the same constraints as in Ref.~\cite{csb}. % [Fig.~\ref{qp}(c)]. 

At finite density [Fig.~\ref{qp}(c)], the quasiparticle velocity is decreased by an amount proportional to the density of other quasiparticles, which in turn is (approximately) proportional to the density of occupied sites; however, the dynamics still consists of ballistically moving quasiparticles.
At high density, the model remains integrable in terms of these quasiparticles. However, one can identify the quasiparticle content of a product state by recasting the model in terms of bonds, as follows~\footnote{D.A. Huse, private communication}: assign a quasiparticle to each $AB$ bond where both spins are up, and assign two quasiparticles to any spin configuration that has the sequence $\downarrow \uparrow \downarrow$. Note that the number of physical up spins fluctuates, though the number of quasiparticles remains conserved, because quasiparticles transiently ``merge'' during a collision [Fig.~\ref{qp}(a-b)]. The quasiparticle structure is explored further in~\cite{suppmat}, by simulating the ``free expansion''~\cite{rigol2005free} of a general initial state.

\emph{OTOCs}.---The clearest evidence that the model remains integrable at high densities comes from studying OTOCs. In a classical system, the OTOC corresponds to the local overlap between two histories with identical initial conditions except for a disturbance at the origin~\cite{das2017light}. 
To establish integrability we perform the following numerical experiment [Fig.~\ref{qp}(c-d)]: we create an initial state with a region where the density of up spins in part of the system is low (so we can reliably create a single quasiparticle there) but the rest of the system is at high density. We then move this quasiparticle, and overlap histories with different initial positions of the quasiparticle.  
We find that translating a single quasiparticle does not have a ``butterfly'' effect: the OTOC is simply the time trace of the quasiparticle that was moved [Fig.~\ref{qp}(d)]. This is direct evidence that the model remains integrable at high densities: in a chaotic system, ``firing'' a quasiparticle into the system at time time $t + \delta t$ rather than at $t$ would cause a butterfly effect, which is clearly absent here.

Although moving quasiparticles does not cause a spreading disturbance, adding or removing quasiparticles does, as the presence of a new quasiparticle modifies the phase shifts of all the others. Physical spin operators typically create and/or move quasiparticles, depending on the underlying product state. In contrast with, e.g., the Ising model, all simple operators have an amplitude for both creating and translating quasiparticles. For instance, $\downarrow \uparrow \downarrow \uparrow \downarrow$ is a state with four quasiparticles, while $\downarrow \uparrow \uparrow \downarrow \downarrow$ is a state with only two quasiparticles. Therefore all simple operators spread.

%Even an operator such as $\sigma^-_i \sigma^-_{i+1} \sigma^+_{i+2} \sigma^+_{i + 3}$, which might naively seem to involve moving a quasiparticle, can end up changing the quasiparticle number (since $\ldots\downarrow \uparrow \uparrow \uparrow \uparrow \downarrow\ldots$ has three quasiparticles [Fig.~\ref{qp}(b)], but $\
%
%: for example, the sequence $\ldots0000111100\ldots$ contains three quasiparticles (Fig.~\ref{qp}(b)), while the state $\ldots 0011001100\ldots$ (related by the operator above) contains two quasiparticles. In a typical eigenstate, every simple operator will change the quasiparticle number in at least some of the spin configurations represented in the eigenstate, and therefore spread.

%specifically the out-of-time-order commutator of two operators, $C(x,t) \equiv \langle [X_0(0), Z_x(t)]^2 \rangle$.

\begin{figure}[tb]
\begin{center}
\begin{minipage}{0.23\textwidth}
\includegraphics[width = \linewidth]{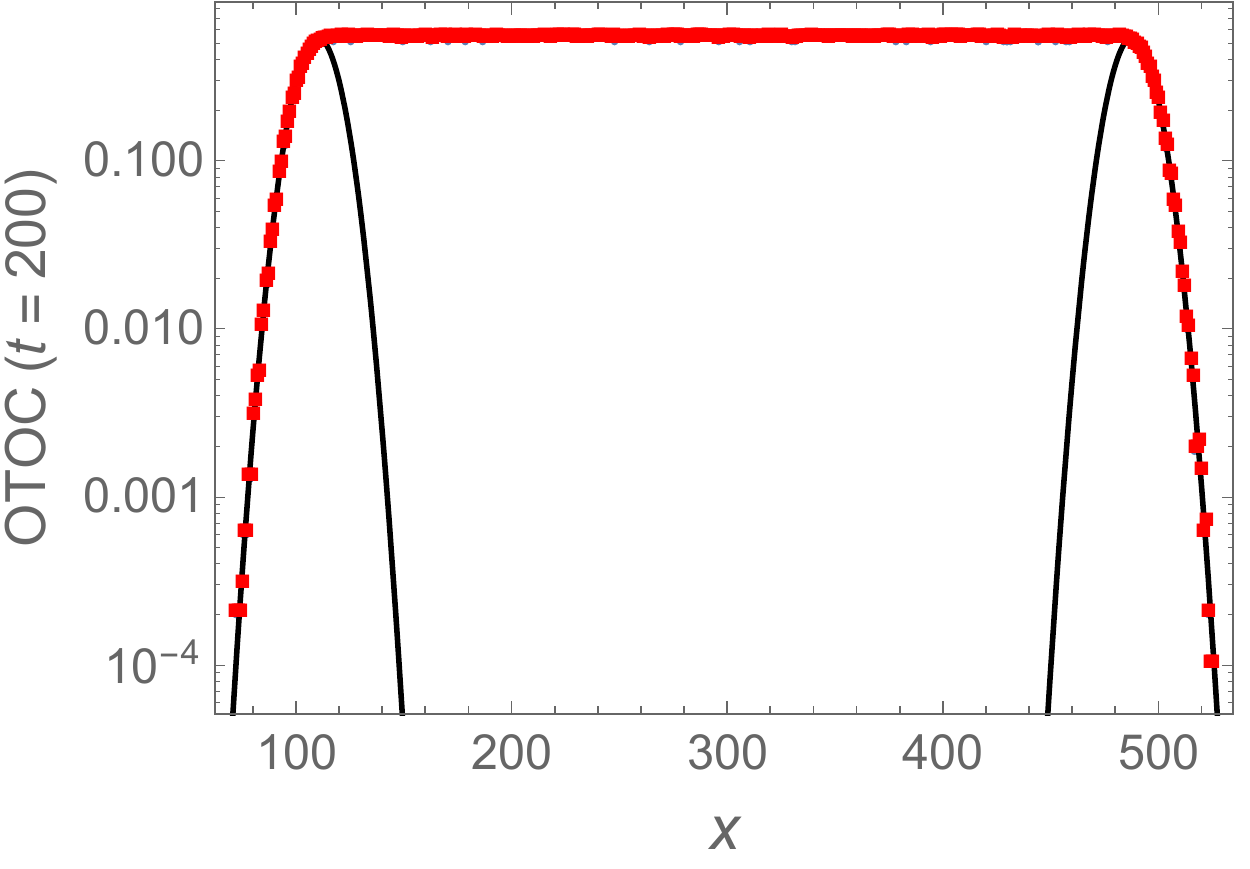}
\end{minipage}
\begin{minipage}{0.23\textwidth}
\includegraphics[width = \linewidth]{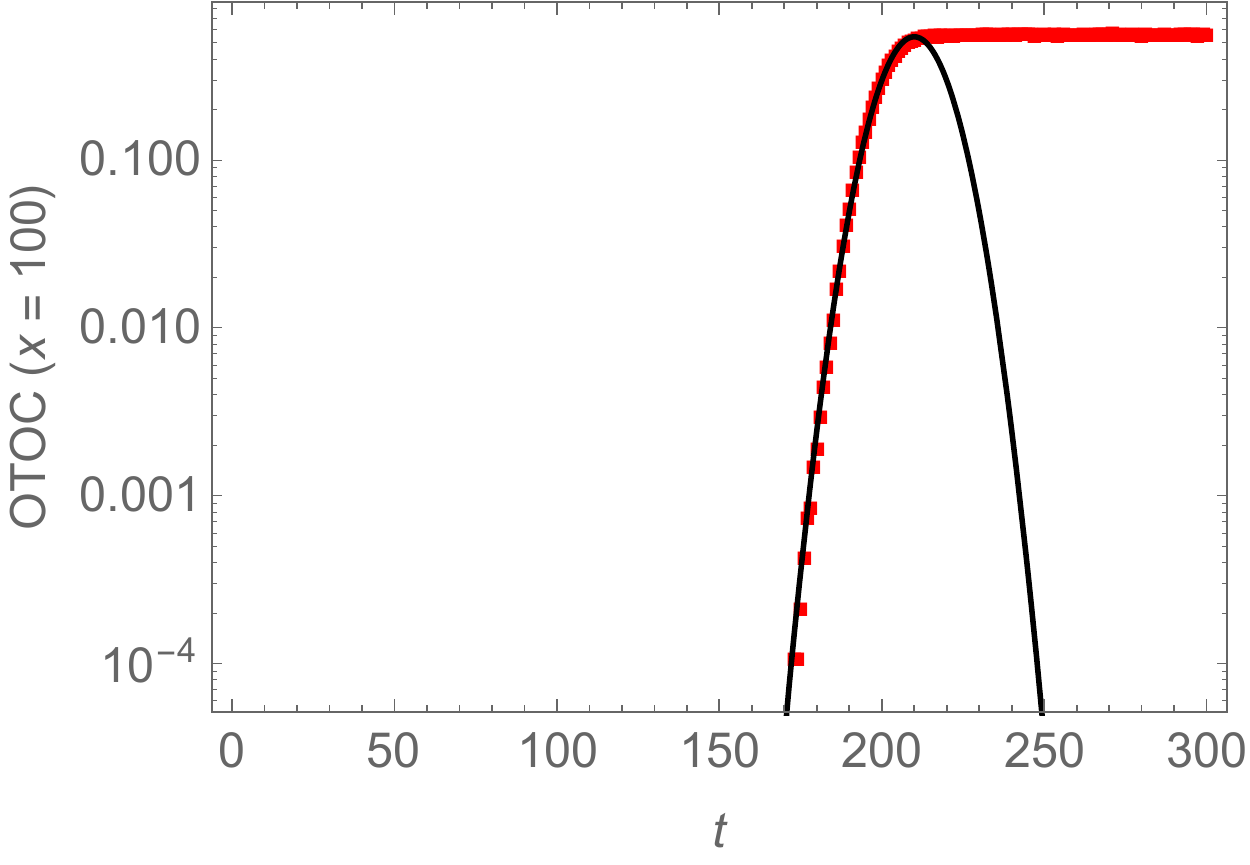}
\end{minipage}
\begin{minipage}{0.23\textwidth}
\includegraphics[width = \linewidth]{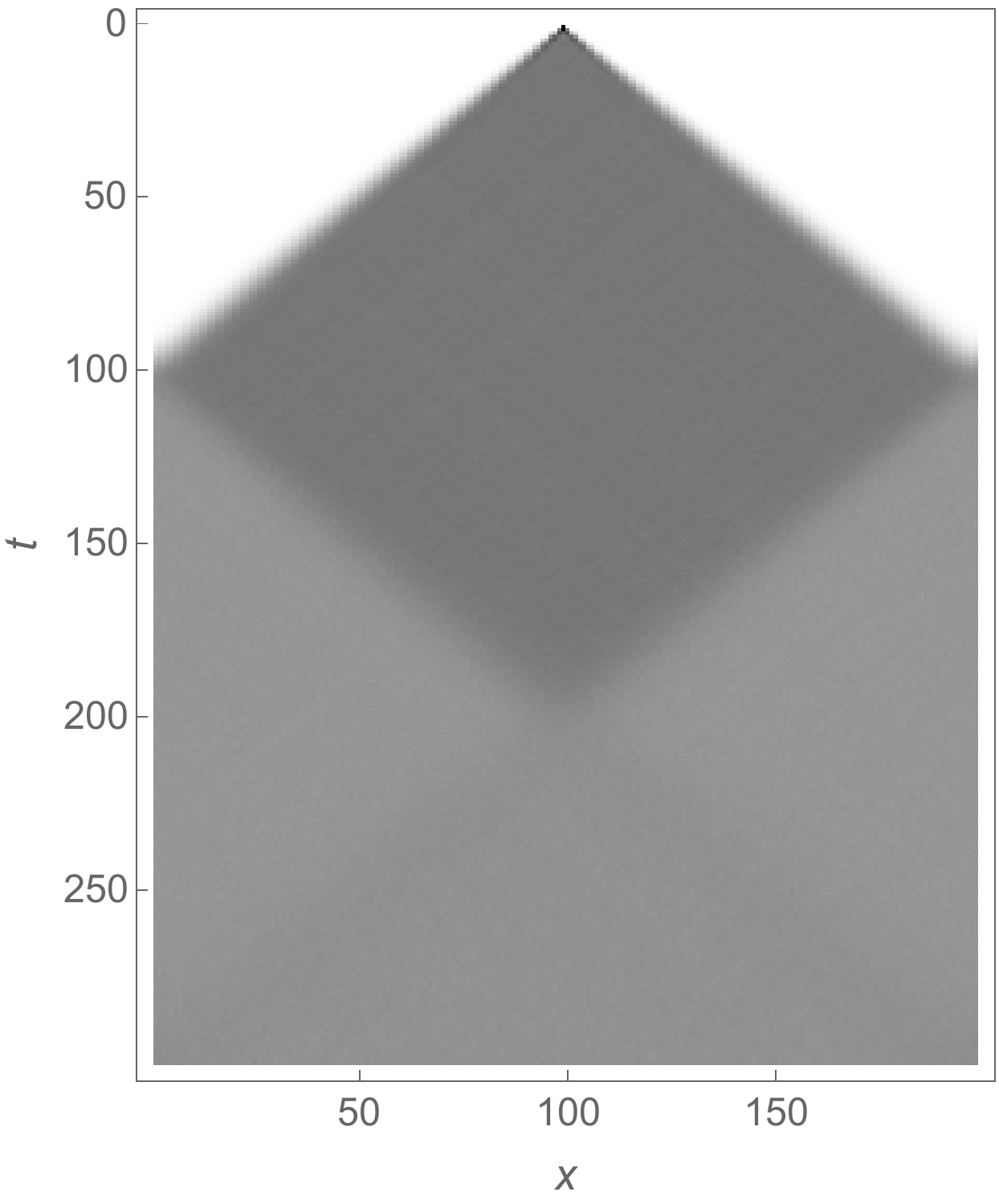}
\end{minipage}
\begin{minipage}{0.23\textwidth}
\includegraphics[width = .95 \linewidth]{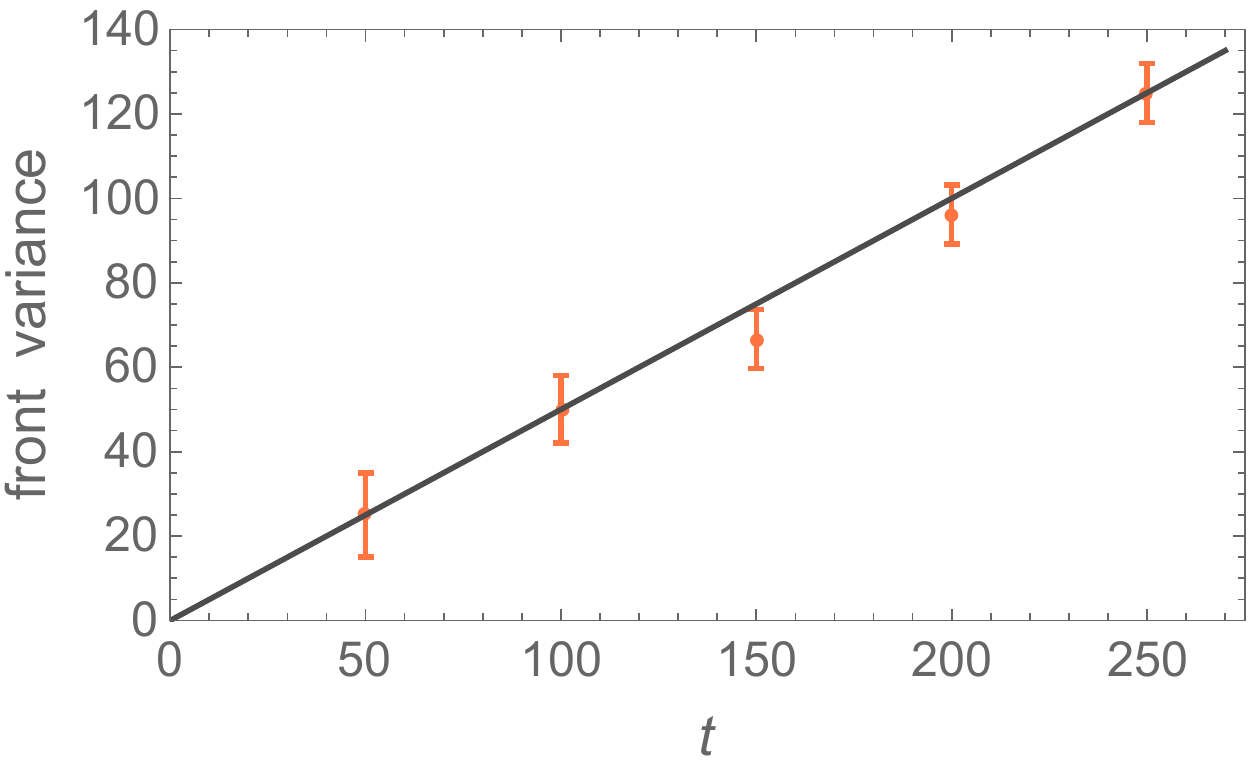}
\includegraphics[width =  .95 \linewidth]{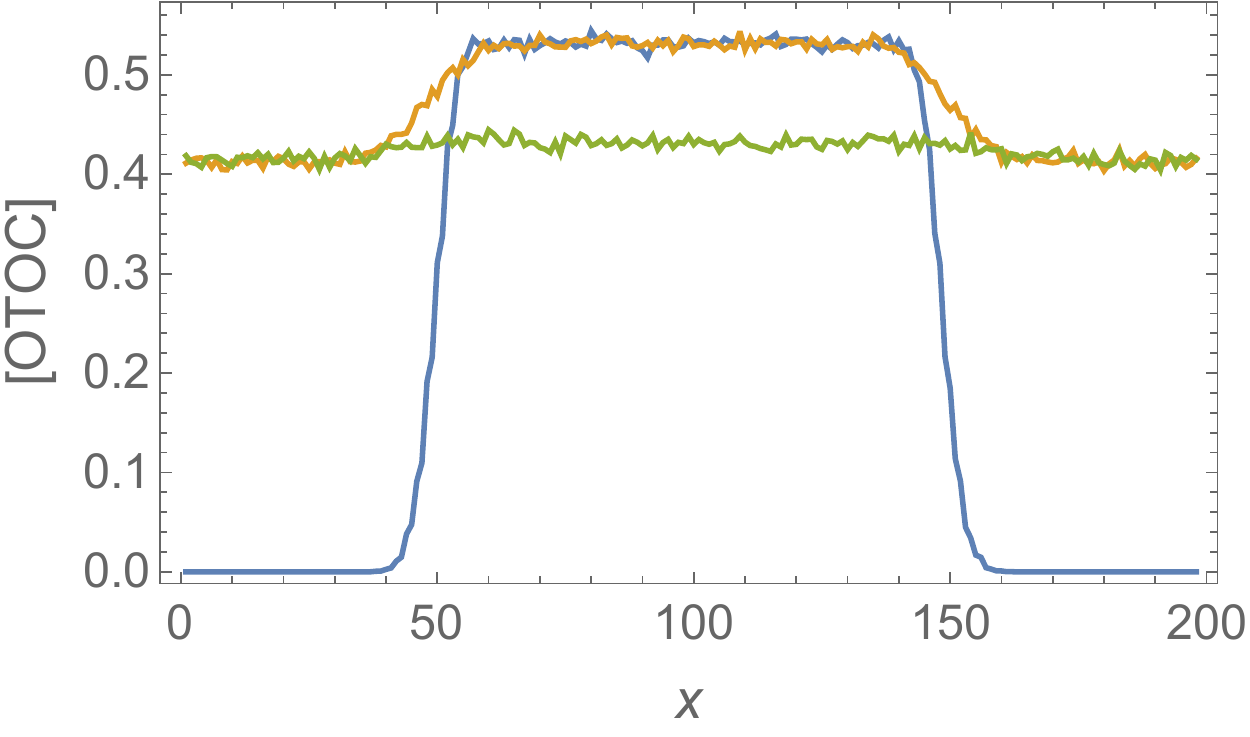}
\end{minipage}
\caption{Upper panel: Value of the out-of-time-order commutator at a fixed time, vs. position (left) and at a fixed position, vs. time (right), for $L = 600$ systems. The operator front is well described by a Gaussian. Middle right: squared width (variance) of operator front vs. time, indicating diffusive broadening. Lower left: density plot of the OTOC vs. time, for a smaller system ($L = 200$), showing the unexpected overshooting effect at times longer than system size; cross-sections at $t = 50, 100, 150$ are shown in the lower right panel.}
\label{fig3}
\end{center}
\end{figure}

For concreteness we focus on the simplest OTOC, $C(x,t) \equiv \langle [X_0(0), Z_x(t)]^2 \rangle$. (We have also checked the OTOCs of more complicated operators~\cite{suppmat} but they do not behave appreciably differently, in contrast with the Ising case~\cite{motrunich_otoc}.) The OTOC averaged over many initial product states grows with a light-cone typical of chaotic systems (Fig.~\ref{fig1}). The magnitude of the OTOC near its growth ``front'' precisely matches the recent prediction for chaotic systems~\cite{xu2018a, xu2018b, khn}, behaving to leading order as $\exp[-(x - vt)^2/(2\sigma^2(t))]$, where $\sigma^2 \sim t$ [Fig.~\ref{fig3}]. The velocity of the OTOC front is half the causal light-cone speed; this is expected, since a random state is at half-filling. The FFA model is thus distinct from other large-$N$ or ``classical'' limits, in which the OTOC front is sharp~\cite{nrvh, xu2018a, xu2018b} (note that in the FFA model the front is sharp \emph{if} the initial state is a computational-basis product state but not otherwise). Starting from a random initial distribution of quasiparticles, the diffusive broadening of the front is a natural consequence of the random time delays due to collisions.

The late-time behavior of $C(x,t)$ has some unusual features, which can be seen in the lower two panels of Fig.~\ref{fig3}. Even at times before the operator has wrapped around the system, its behavior inside the front is unexpected. Rather than decorrelating completely, the two histories remain weakly \emph{anticorrelated} within the front (so the value of the OTOC inside the front is $\approx 0.54$ rather than $1/2$). After a timescale $t \sim L/2$, this behavior changes and the two histories compared by the OTOC become weakly \emph{correlated}, giving rise to the diamond-like shape seen in the space-time plot [Fig.~\ref{fig3}, lower left]. The OTOC then saturates at a value $\approx 0.42$ until the much longer revival timescale $T_r$. These saturation values vary depending on the operator. These effects stem from high-density initial configurations~\cite{suppmat}. 

%We have not systematically investigated the outlying states that violate ETH at large system sizes. However, there are 

Our quantitative analysis of the OTOC involved averaging over eigenstates. The OTOC in a single randomly chosen eigenstate also spreads out with a light-cone, but there is less broadening, and the front ``refocuses'' on the timescale $T \sim L/2$~\cite{suppmat}. This can be understood within the quasiparticle picture. In a particular eigenstate, the distribution of left and right moving quasiparticles, and the spacings among the left- and right-movers, are fixed, but one averages over the point in the classical trajectory at which the new quasiparticle is introduced. Thus, the sequence of time delays experienced by (say) a right-mover is randomized, but the total time delay is fixed by the total number of left-movers. Nevertheless, we expect the broadening of the front to be Gaussian at times $t \ll L$, since in this temporal regime the average runs over randomly timed collisions.

\emph{Eigenstates}.---We now turn to the eigenstates of the FFA model. Since $U$ takes each product state to a product state, the dynamics of an initial product state consists of chains of transitions $|C_1 \rangle \mapsto |C_2 \rangle \mapsto |C_3 \rangle \ldots \mapsto |C_N \rangle \mapsto |C_1 \rangle \mapsto |C_2 \rangle \ldots$. A random eigenstate can therefore be constructed~\cite{gz} by picking a random initial product state and summing over its orbit, with appropriate phases, i.e., 

\beq
|E\rangle = \frac{1}{\sqrt{N}} (|C_1\rangle + e^{i q} |C_2\rangle + \ldots + e^{(N - 1) i q} |C_N \rangle).
\eeq
Such states are evidently eigenstates of $U$ so long as $Nq = 2\pi n$ for some $n$. %One can generate a random eigenstate by starting with a random product state, time-evolving it until recurrence, and summing over its orbit~\cite{gz}. 

\begin{figure}[tb]
\begin{center}
\begin{minipage}{0.23\textwidth}
\includegraphics[width = \linewidth]{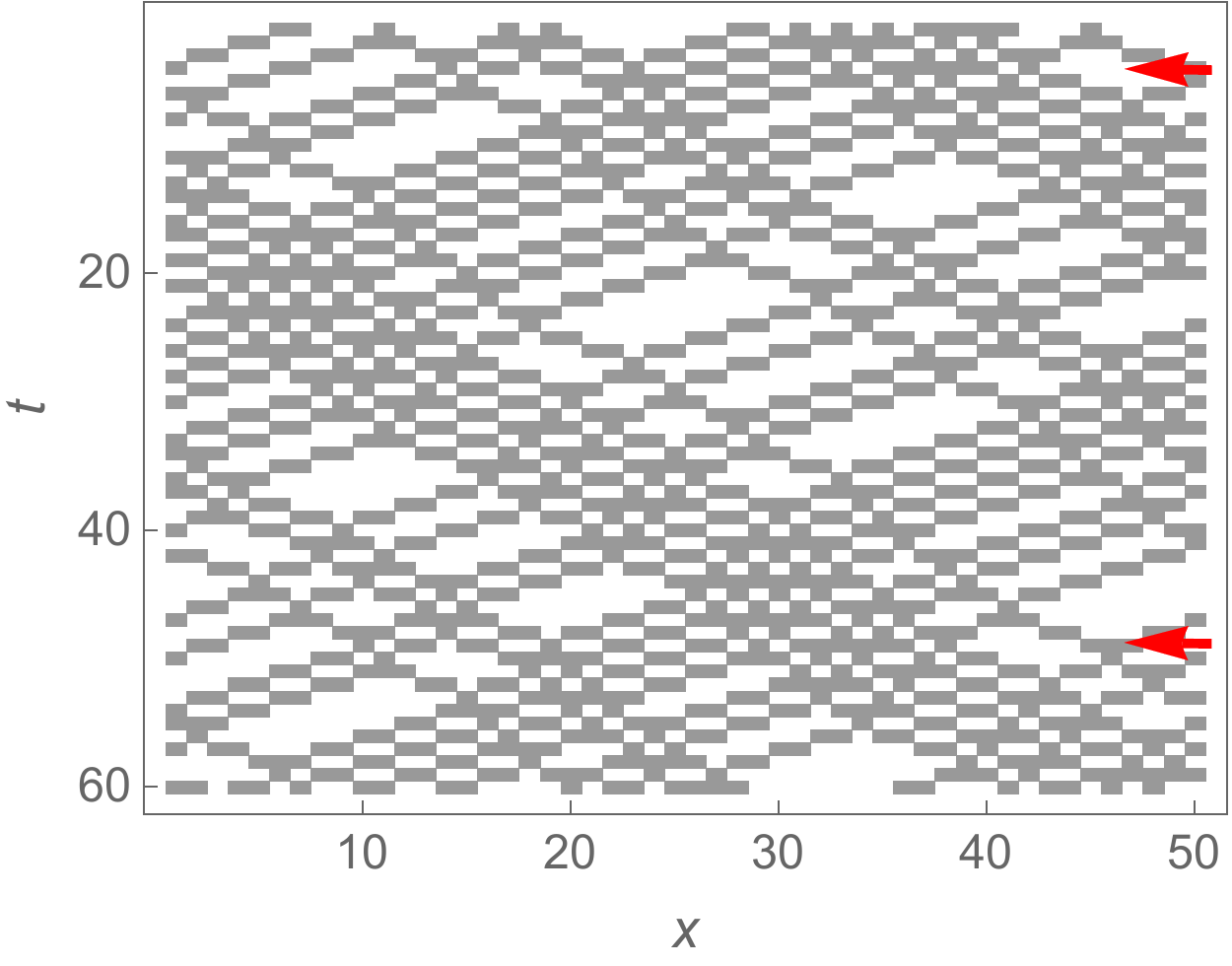}
\end{minipage}
\begin{minipage}{0.23\textwidth}
\includegraphics[width = \linewidth]{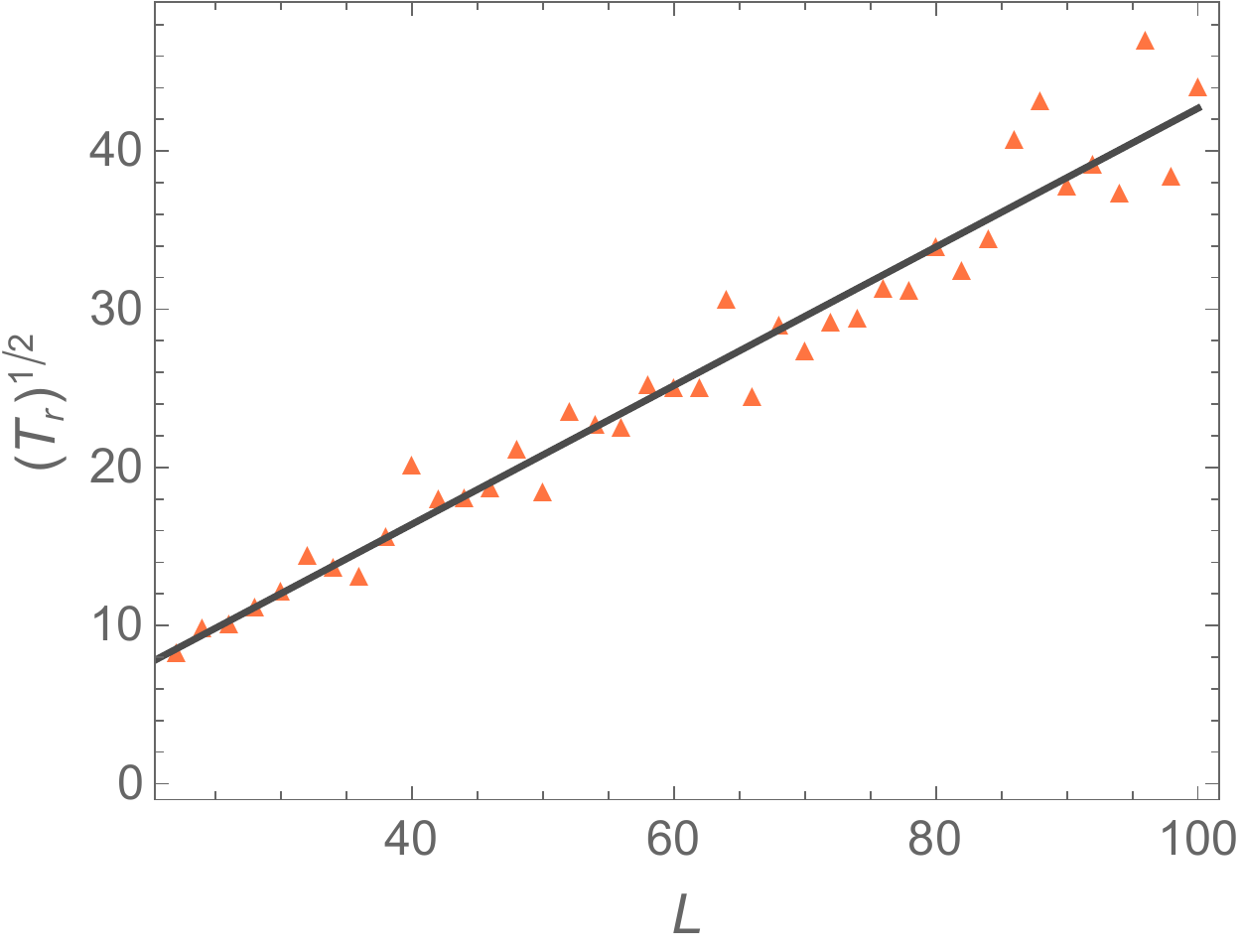}
\end{minipage}
\begin{minipage}{0.23\textwidth}
\includegraphics[width = \linewidth]{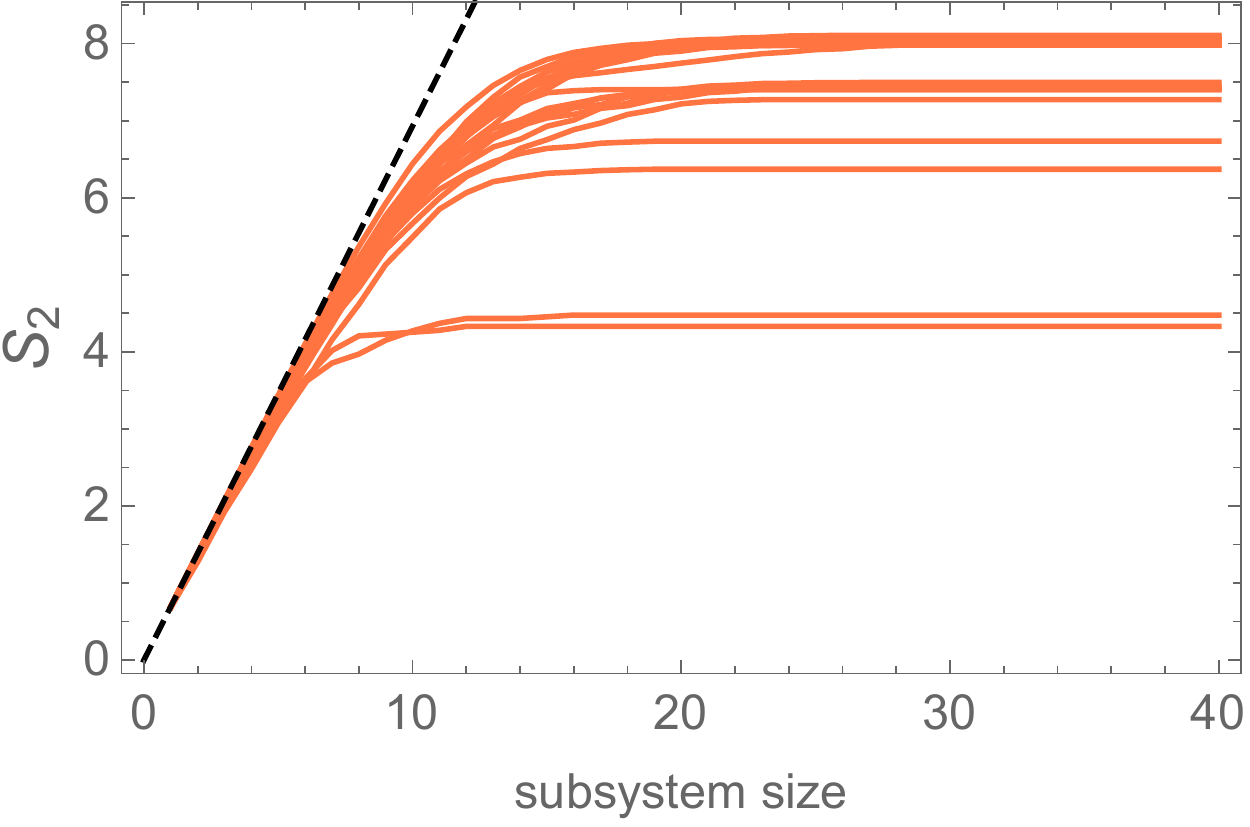}
\end{minipage}
\begin{minipage}{0.23\textwidth}
\includegraphics[width =  \linewidth]{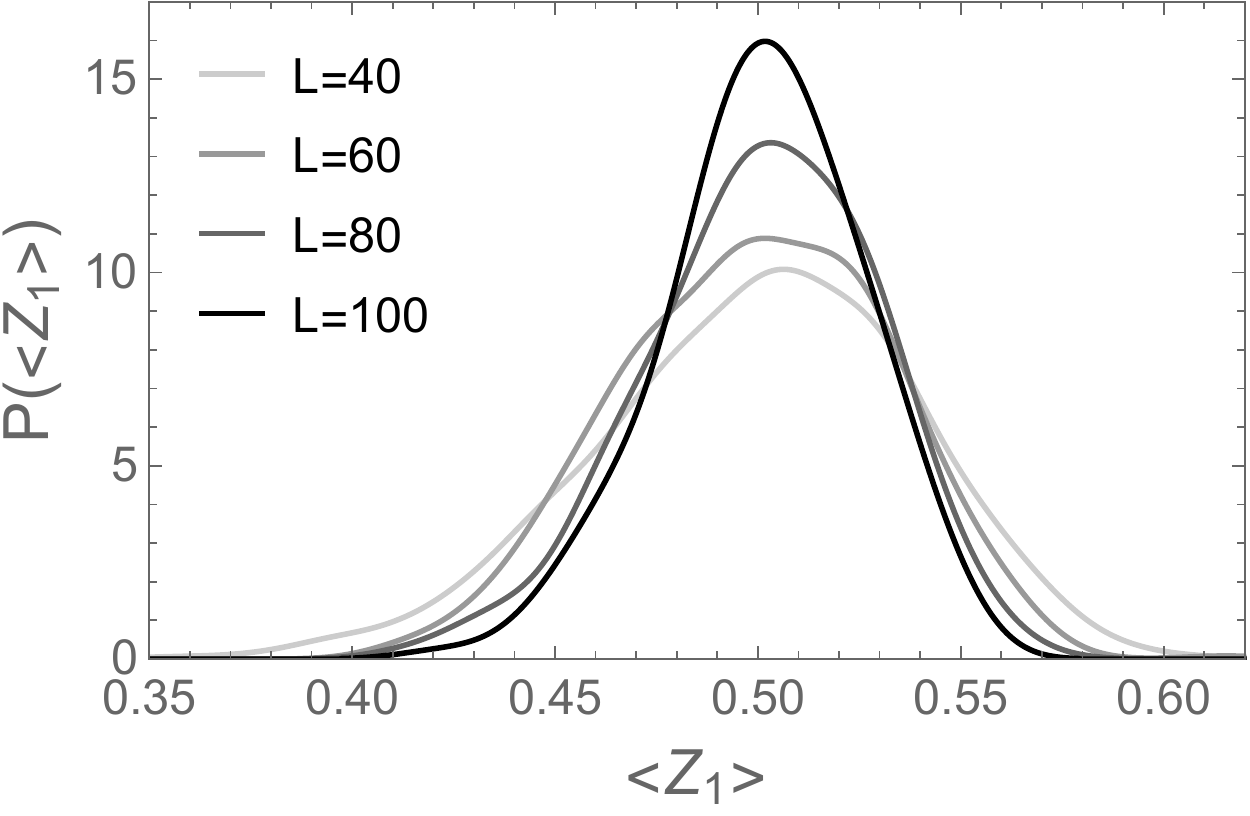}
\end{minipage}
\caption{Upper left: An example of the dynamics of an initial product state, illustrating how configurations recur after a period of order $L$ up to an overall translation (see arrows; in this case the period is 45 and the shift is four sites to the left). Upper right: typical $\sqrt{T_r}$ vs. system size, as predicted by the quasiparticle picture. 
Left: State-to-state fluctuations of Renyi entropy $S_2$, vs. subsystem size, for $L = 80$; each line represents a randomly chosen eigenstate. Right: histogram of the expectation value of the spin on the first site, $\langle Z_1 \rangle$, across eigenstates, for various system sizes. The histograms narrow with increasing system size.}
\label{fig2}
\end{center}
\end{figure}

A central quantity in our analysis is the recurrence time $T_r$ for a given initial state, which measures how much of configuration space is accessible under unitary evolution from a given initial state. In an ``ergodic'' system, essentially every configuration would be visited, and $T_r$ would grow exponentially with system size. This is not what happens in the FFA model (Fig.~\ref{fig2}); instead $T_r \sim L^2$. This follows from the quasiparticle picture: a left-mover traverses the system on a timescale set by the number of right-movers, and vice versa. Since both numbers are of order $L$, their least common multiple is $\sim L^2$, although there are many configurations with a much smaller least common multiple, and for these $T_r$ is much smaller (Fig.~\ref{outliers}). 
%$S_p$ appears to grow linearly with system size, but with a much reduced coefficient $0.03 \ll \ln 2$. Thus, the eigenstates of this model are not ergodic, but ``fractal'' in the sense that their size scales as an anomalous small power of the Hilbert space dimension. 
Note that $T_r$ is a recurrence time specific to a particular initial computational-basis product state. The recurrence time of a random initial vector $T_g$ is the least common multiple of each orbit's recurrence time. By sampling many initial states and computing their $T_r$, we find that this global recurrence time grows at least as fast as $T_g \agt 4.8^L$~\cite{suppmat}, so it is in general exponentially larger than the Hilbert space dimension. The quasiparticle picture also suggests that the scaling is exponential with system size: according to this picture, $T_g$ is the least common multiple of all possible quasiparticle periods $\alt L$, which scales as the product of all primes $\alt L$; asymptotically this product grows exponentially in $L$ by the prime number theorem.
% can also see that $T_g \leq (L!)^2$, since this is a lower bound on the least common multiple of all possible quasiparticle periods. 
Thus, $T_g$ scales exponentially rather than double-exponentially with system size, in contrast with generic chaotic models~\cite{hosur2016chaos}.

% (though likely smaller than the double exponential scaling expected for generic chaotic Hamiltonians). 

We previously showed~\cite{gz} that $\ln T_r$ upper-bounds the second Renyi entanglement entropy $S_2$~\cite{nielsen_chuang} through any bipartite cut; an implication is that $S_2(\ell) \alt 2 \ln L$ for any subsystem size $\ell$ and system size $L$. This is what we find, by directly computing $S_2$ for various subsystem sizes (Fig.~\ref{fig1}). As Fig.~\ref{fig2} shows, there are strong state-to-state fluctuations in the saturation value of $S_2$, but the entanglement of every eigenstate saturates for subsystems well below half the system size. Small subsystems, on the other hand, appear to satisfy ETH: both the Renyi entropy for small subsystems of length $\ell \alt 5$ and the expectation values of on-site operators are narrowly distributed, with state-to-state spread that narrows with system size (Fig.~\ref{fig2}). This narrowing is slower than for generic ETH systems, however. More details about the outlying states are discussed in~\cite{suppmat}.

%\begin{figure}[tb]
%\begin{center}
%\begin{minipage}{0.23\textwidth}
%\includegraphics[width = .95 \linewidth]{histog2b}
%\end{minipage}
%\begin{minipage}{0.23\textwidth}
%\includegraphics[width =  .95 \linewidth]{outlier_frac}
%\end{minipage}
%\caption{Left: histogram of $T_r/L$, showing the outlier peak at $T_r \approx L$. Right: fraction of outliers vs. system size.}
%\label{outliers}
%\end{center}
%\end{figure}

%We now turn to outlier states that are maximally nonthermal~\cite{ETH_outliers}; our proxy for non-thermality is an anomalously low $T_r$. Fig.~\ref{outliers} histograms the distribution of $T_r$ across eigenstates at a given system size; note the ``outlier'' peak at $T_r \approx L$.  Exponentially many such outliers can be constructed, if one begins with states that have an atypical quasiparticle distribution. What is less certain is whether these outliers constitute a finite fraction of all states in the thermodynamic limit. Fig.~\ref{outliers} histograms the distribution of $T_r$ across eigenstates at a given system size; there is an ``outlier'' peak at $T_r \approx L$. 
%%
%Since our numerical approach is restricted by the growth of $T_r$, we can track the weight of this outlier peak for quite large systems (up to $L \approx 2000$ sites). As system size is increased, the outliers rapidly drop to about $5\%$ of the states. At large system sizes, the outlier ``density'' apparently decreases further with increasing system size (Fig.~\ref{outliers}), but we have not been able to identify the functional form for this decrease. 

\emph{Eigenvalues and level statistics}.---From the construction of eigenstates, it follows that each eigenvalue is of the form $\omega_n = 2\pi m /T_r^{(n)}, 0 \leq m \leq 2\pi T_r^{(n)}$, where $n$ labels inequivalent orbits. There is an $n$-fold degeneracy at quasienergy zero (i.e., eigenvalue unity for the unitary $U$), and also other large degeneracies at special frequencies that divide many orbit periods. Owing to these degeneracies the level statistics are neither Poisson nor random-matrix. %(However, it is possible that trivial modifications such as adding phase gates that do not affect the eigenstates will suffice to restore Poisson statistics.)

\begin{figure}[bt]
\begin{center}
\begin{minipage}{0.23\textwidth}
\includegraphics[width = 0.93 \linewidth]{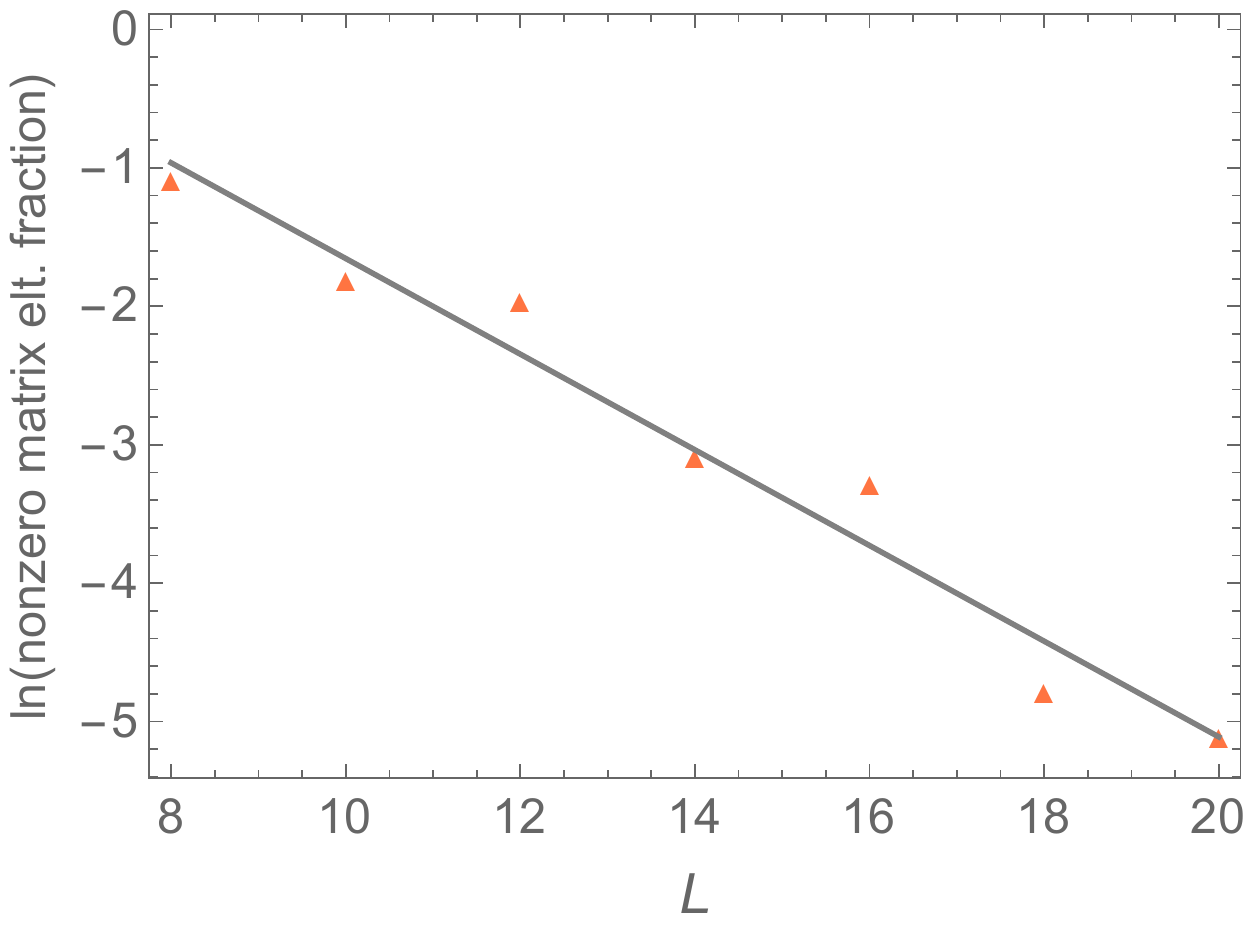}
\end{minipage}
\begin{minipage}{0.22\textwidth}
\includegraphics[width =  \linewidth]{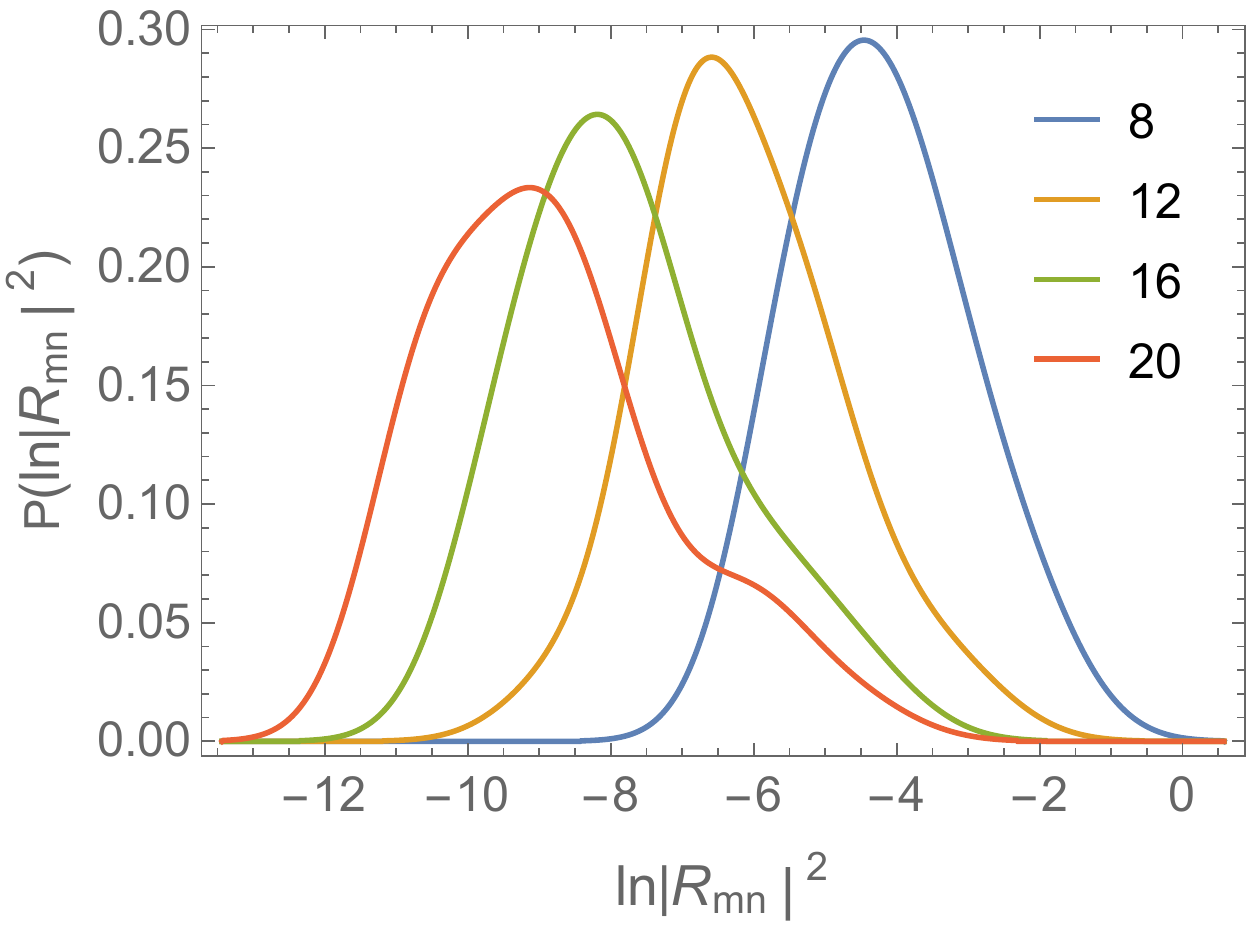}
\end{minipage}
\caption{Left: the fraction of nonzero off-diagonal matrix elements falls off exponentially with system size; the slope of the fit line is $-\ln 2/2$. Right: the distributions of the remaining nonzero matrix elements broaden (histograms are rescaled to have the same weight).}
\label{odm}
\end{center}
\end{figure}

\emph{Off-diagonal matrix elements}.---Finally, we consider the off-diagonal matrix elements of local operators (specifically, the two-spin-flip operator $\sigma^x_i \sigma^x_{i+1}$) between eigenstates. For convenience we restrict ourselves to matrix elements between states in the quasi-energy zero sector; as noted above, the spectrum is highly degenerate and each distinct ``orbit'' contributes one state to this sector. Most pairs of eigenstates have strictly zero matrix element; the fraction of nonzero matrix elements decreases exponentially with system size, approximately as $2^{-L/2}$ (Fig.~\ref{odm}). This behavior has also been seen in other integrable models~\footnote{V. Khemani, private communication.}. The distribution of the nonzero matrix elements broadens with system size but does not seem to approach a Gaussian at the accessible sizes.%; however, that the system sizes we have been able to explore are relatively small. (The method employed here is to construct eigenstates by random sampling, so we are limited by the rapidly increasing fraction of zero matrix elements, as the number of samples required to find a reliable number of nonzero matrix elements grows exponentially.)

\emph{Discussion}.---This work analyzed the FFA model, a simple interacting integrable Floquet system. Because this model has a special basis with classical dynamics, various quantities can be computed here that are harder to compute in more realistic interacting integrable systems such as the Heisenberg chain. In its integrable dynamics the FFA model is analogous to a discretized hard-rod gas (except that collisions here delay the propagation of quasiparticles instead of speeding it up). Despite the integrability of the model, physical operators exhibit a \emph{chaotic} butterfly effect, with a diffusively broadening front that quantitatively matches the predictions of Ref.~\cite{xu2018a}. We believe this feature should be generic for interacting integrable systems, as it arises from the medium-dependence of the velocities, which is a generic feature.
Moreover, small subsystems (relative to the system size) obey the eigenstate thermalization hypothesis. Thus, integrability is ``hidden'' from these diagnostics. However, large subsystems clearly violate the ETH: the eigenstate entanglement has a sharpening crossover from volume-law growth to saturation, at a subsystem size that grows logarithmically with the full system size. Metrics such as the level statistics also diagnose the non-thermal character of this model. %Unfortunately, the more physically reasonable measures are also the least sensitive to integrability. 

Many questions remain for future work, including perturbations of the model that restore chaos and/or quantum fluctuations; identifying the local conserved operators and formally demonstrating the integrability of the model, e.g., through methods for integrable cellular automata~\cite{ica0, ica1, ica2}; and exploring operator entanglement. 

\emph{Note added}.---While this manuscript was being prepared, a preprint appeared~\cite{rowlands_lamacraft}, mapping operator spreading in noisy quantum circuits to the (stochastic) Frederickson-Andersen model. The results seem conceptually unrelated to ours, however.

\emph{Acknowledgments}.---S.G. thanks David Huse, Vedika Khemani, Brian Swingle, and Romain Vasseur for helpful discussions and comments on a draft of this paper, and Lincoln Carr, Andrew Potter, and Bahti Zakirov for helpful discussions on related topics. This work was supported by NSF Grant No. DMR-1653271.



%\bibliography{kcm}

\begin{widetext}

\section*{Supplemental Material}

In what follows, we discuss the quasiparticle structure of the FFA model at general densities, quantify states that are outliers in their entanglement properties, estimate of the global recurrence time, and present numerical results on the OTOCs of more complicated operators as well as OTOCs measured within single product states or single eigenstates. 

\emph{Structure and counting of quasiparticles}.---To extract the quasiparticle content of a general initial state, we perform the following numerical experiment. We create an initial state with no up spins outside of a region of size $L$, and observe its time dynamics. The initial state is composed of ballistic quasiparticles, and under time evolution these fly apart; at timescales $t \agt L$, the left- and right-movers will have completely separated, allowing us to resolve the quasiparticle content of the state. Fig.~\ref{expansion} shows this expansion dynamics for a random initial state, as well as a very high-density initial state in which all spins are up. 

\begin{figure}[htbp]
\begin{center}
\includegraphics[width = 0.5\textwidth]{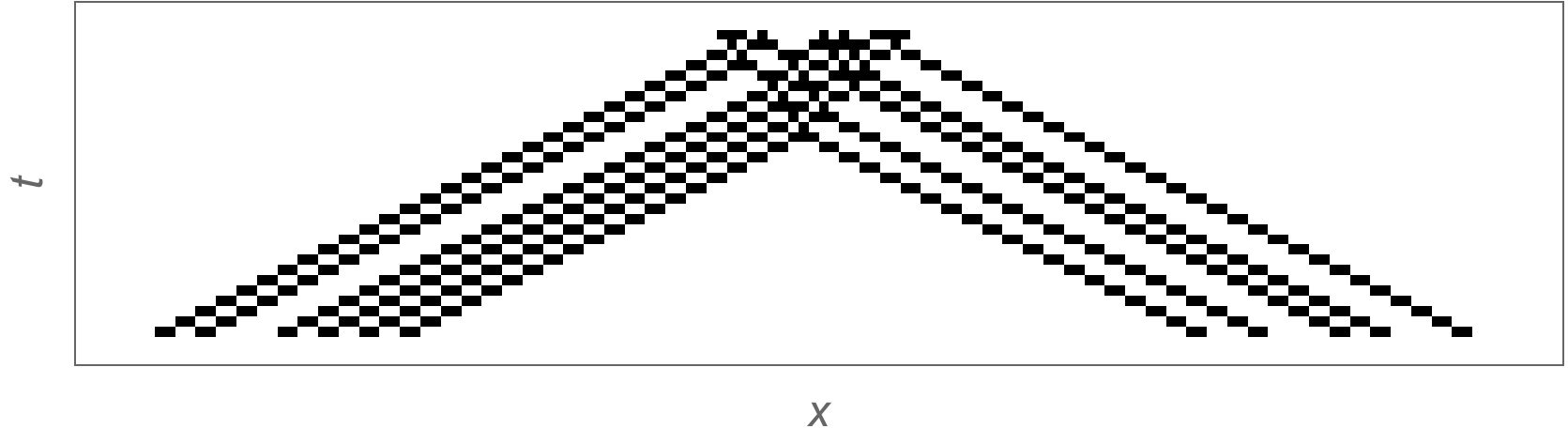}
\includegraphics[width = 0.5\textwidth]{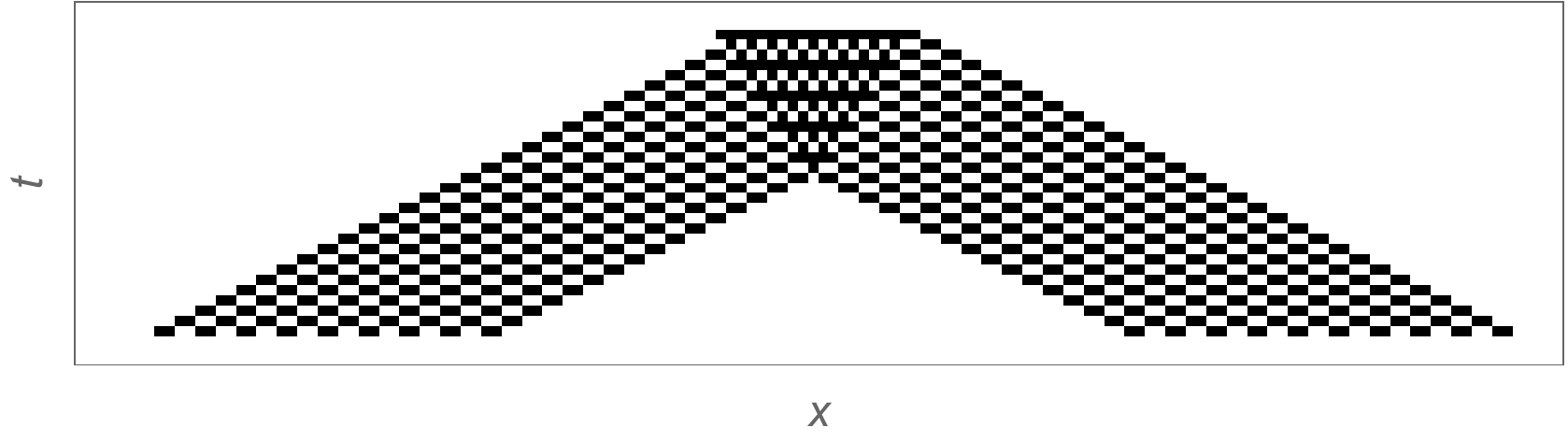}
\includegraphics[width = 0.5\textwidth]{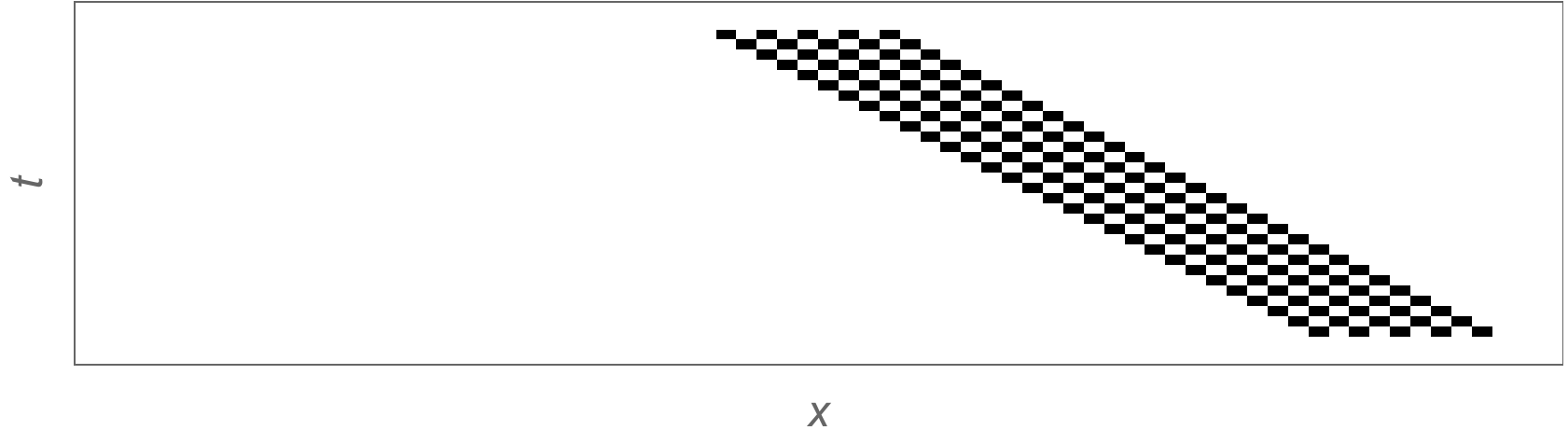}
\caption{Expansion dynamics of initial states that are (respectively) random, all-occupied, and made up only of right-movers.}
\label{expansion}
\end{center}
\end{figure}

These results suggest the following picture for the quasiparticle content of a state, and for entropy vs. number of quasiparticles. (For specificity we consider right-movers; all statements in the rest of this paragraph hold under exchanging right- and left-movers.) First, in general, quasiparticle states can be specified by the \emph{asymptotic} spacings between adjacent right-movers. Second, two right-movers cannot occupy adjacent states; there is a hardcore constraint, as noted in the main text. Third, the number of right-movers that can fit in a system of size $L$ is a function of the number of left-movers, because left-movers effectively ``shrink'' the right movers and allow them to fit closer together. In a system of size $L$, the number of available right-moving states is $L/2 + \text{Floor}(N_l)/2$ where $N_l$ is the number of left-movers. If we guess that the count of left-movers is $L/2$, the average number of right-movers in the state is $1/3 \times 3/4L = L/2$ (using results from the Fibonacci chain~\cite{csb}), so this guess is indeed self-consistent.

\emph{OTOC at very high density}.---The diamond-shaped pattern in the OTOC is in large part a result of initial states that are at very high densities. As Fig.~\ref{highdens} shows, a state with all spins initially up undergoes a ``blinking'' pattern in its time evolution; flipping a single spin creates two propagating domain walls in the blinking pattern, which lead to anticorrelation (rather than just decorrelation) in the OTOC. For a state with a high density of up spins, the OTOC remains strongly anticorrelated until the domain wall wraps around the system, and then decorrelates. 

\begin{figure}[htb]
\begin{center}
\begin{minipage}{0.23\textwidth}
\includegraphics[width = .8 \linewidth]{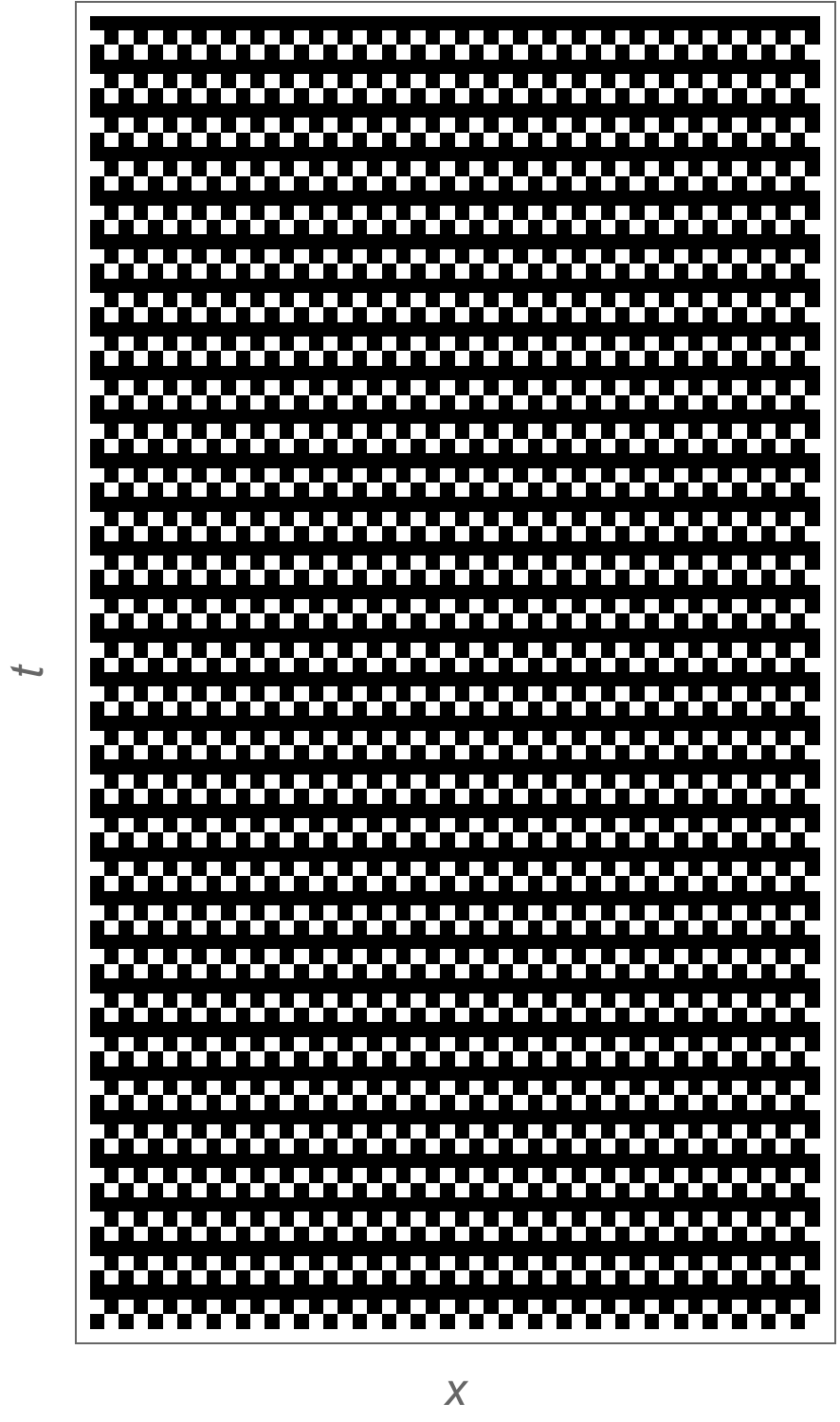}
\end{minipage}
\begin{minipage}{0.23\textwidth}
\includegraphics[width =  .8 \linewidth]{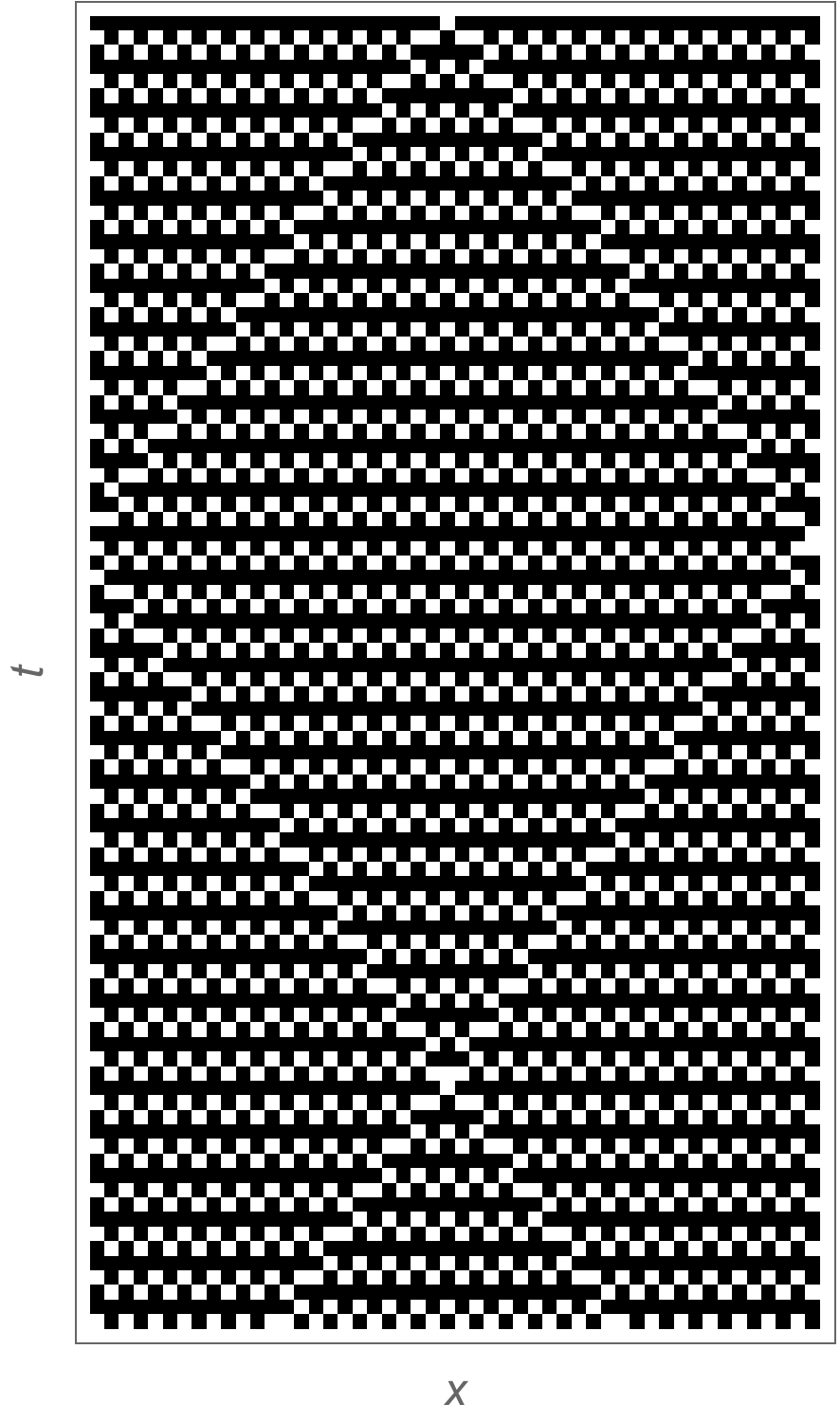}
\end{minipage}
\begin{minipage}{0.23\textwidth}
\includegraphics[width = .8 \linewidth]{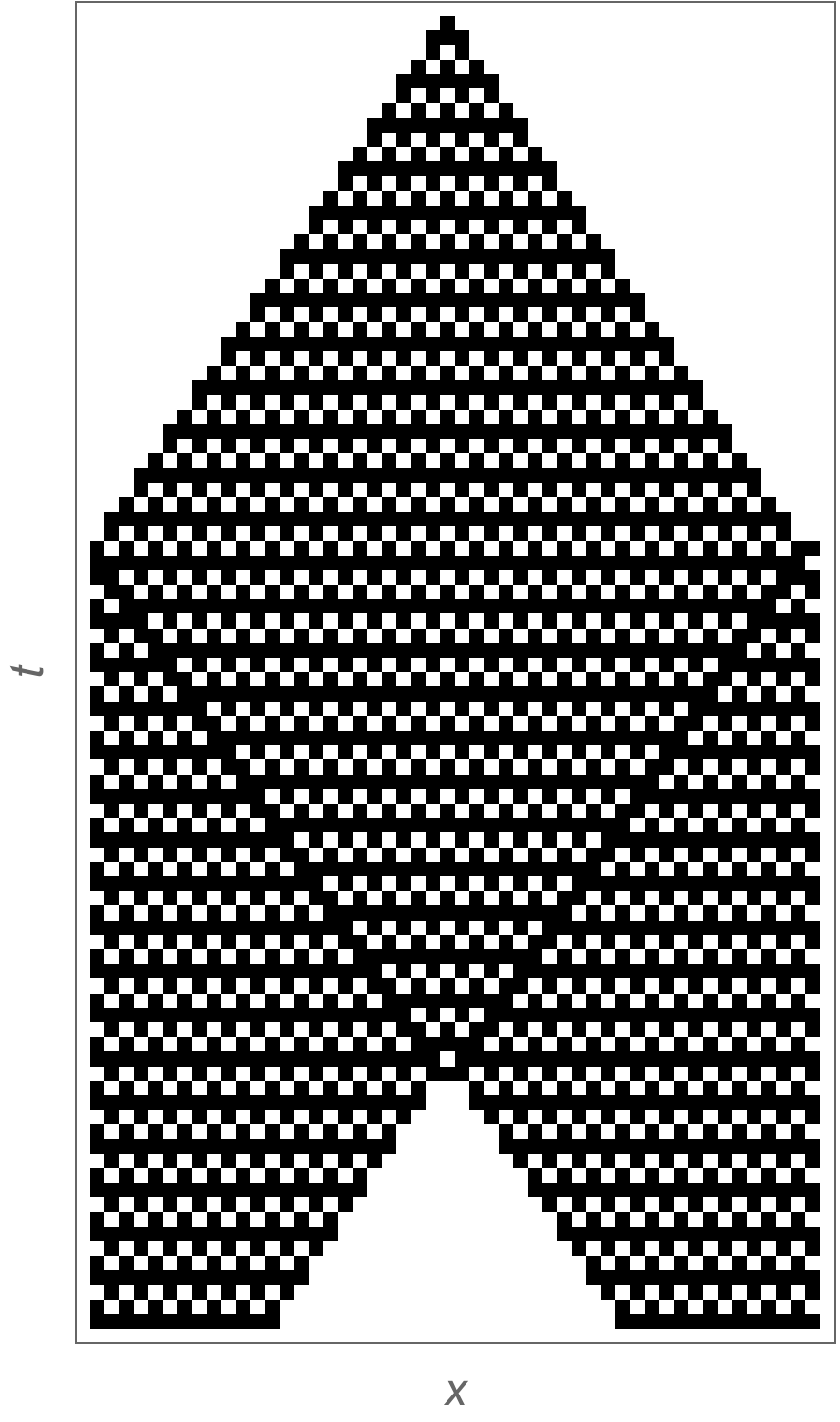}
\end{minipage}
\begin{minipage}{0.23\textwidth}
\includegraphics[width =  .8 \linewidth]{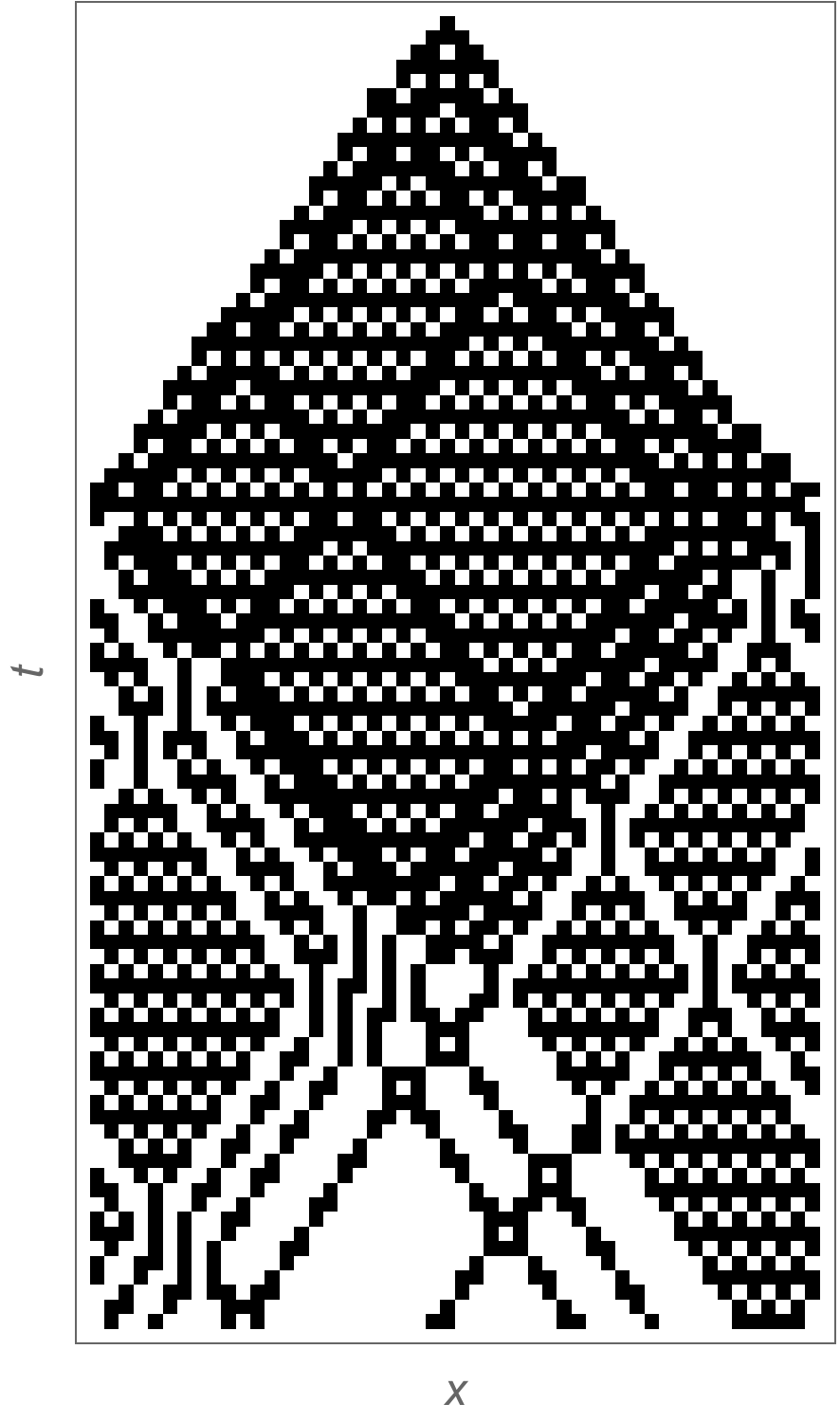}
\end{minipage}
\caption{From left to right: Spacetime plots of a state with all spins initially up, and with a single spin down; OTOC for a state with all spins initially up, and OTOC for a state with a low density of down spins. }
\label{highdens}
\end{center}
\end{figure}

\emph{Outlier states}.---We now turn to outlier states that are maximally nonthermal~\cite{ETH_outliers}; our proxy for non-thermality is an anomalously low $T_r$. Fig.~\ref{outliers} histograms the distribution of $T_r$ across eigenstates at a given system size; note the ``outlier'' peak at $T_r \approx L$.  Exponentially many such outliers can be constructed, if one begins with states that have an atypical quasiparticle distribution. What is less certain is whether these outliers constitute a finite fraction of all states in the thermodynamic limit. Fig.~\ref{outliers} histograms the distribution of $T_r$ across eigenstates at a given system size; there is an ``outlier'' peak at $T_r \approx L$. 
Since our numerical approach is restricted by the growth of $T_r$, we can track the weight of this outlier peak for quite large systems (up to $L \approx 2000$ sites). As system size is increased, the outliers rapidly drop to about $5\%$ of the states. At large system sizes, the outlier ``density'' apparently decreases further with increasing system size (Fig.~\ref{outliers}), but we have not been able to identify the functional form for this decrease. 

\begin{figure}[htb]
\begin{center}
\begin{minipage}{0.23\textwidth}
\includegraphics[width = .95 \linewidth]{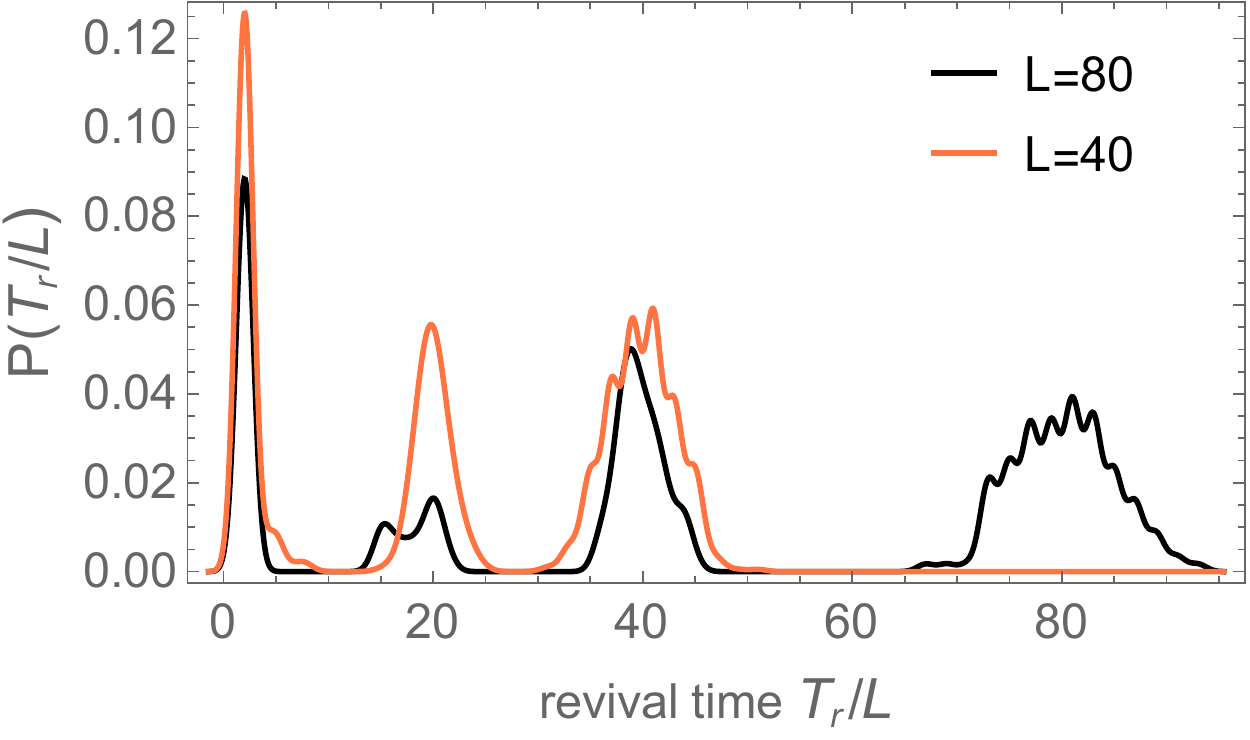}
\end{minipage}
\begin{minipage}{0.23\textwidth}
\includegraphics[width =  .95 \linewidth]{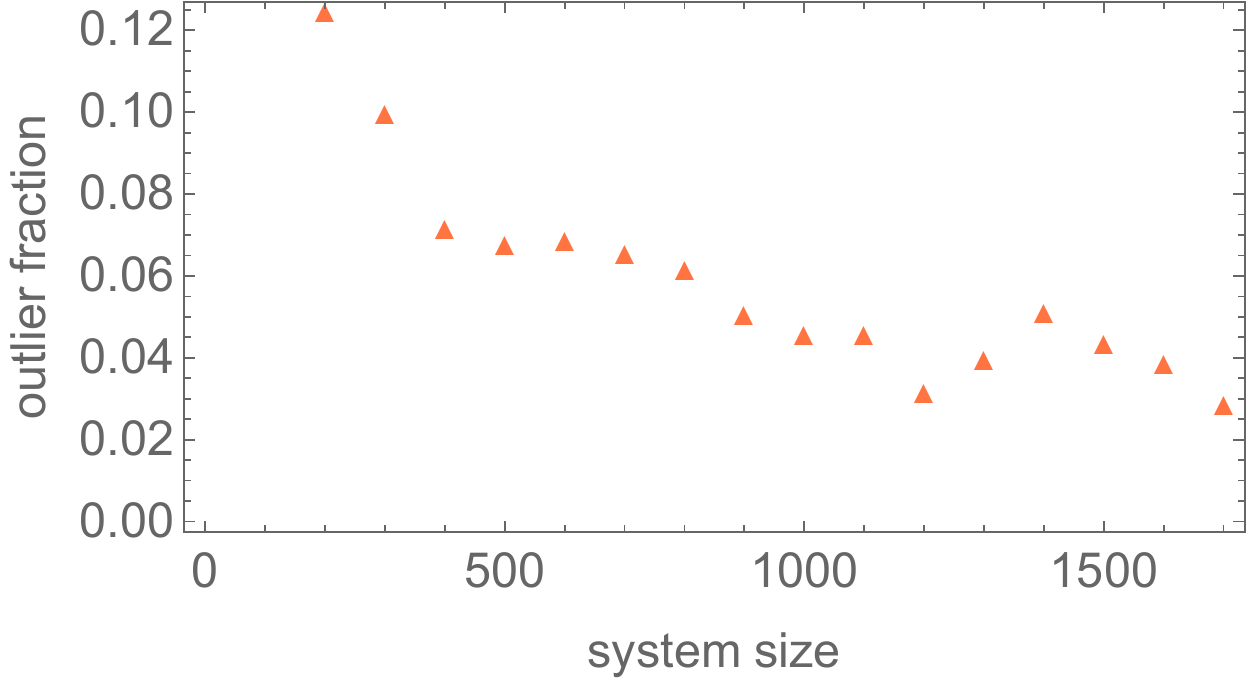}
\end{minipage}
\caption{Left: histogram of $T_r/L$, showing the outlier peak at $T_r \approx L$. Right: fraction of outliers vs. system size.}
\label{outliers}
\end{center}
\end{figure}

\emph{Global recurrence times}.---Our approach to estimating global recurrence times is simply to compute $T_r$ for many random initial configurations (typically $5000$) and then find the least common multiple of all the $T_r$. This is guaranteed to be a lower bound for the global recurrence timescale. This numerical lower bound grows with system size as $4.79^L$ (Fig.~\ref{Tg}). For the smallest system sizes $L = 8, 10$ we have checked explicitly against exact diagonalization that our procedure gives the correct global recurrence timescale. We expect it to be reasonably accurate for the system sizes we have studied, as sampling over 2500 initial conditions instead of 5000 never misses more than one or two frequencies (so our estimate is probably off by only a few orders of magnitude). 

\begin{figure}[htbp]
\begin{center}
\includegraphics[width = 0.3\textwidth]{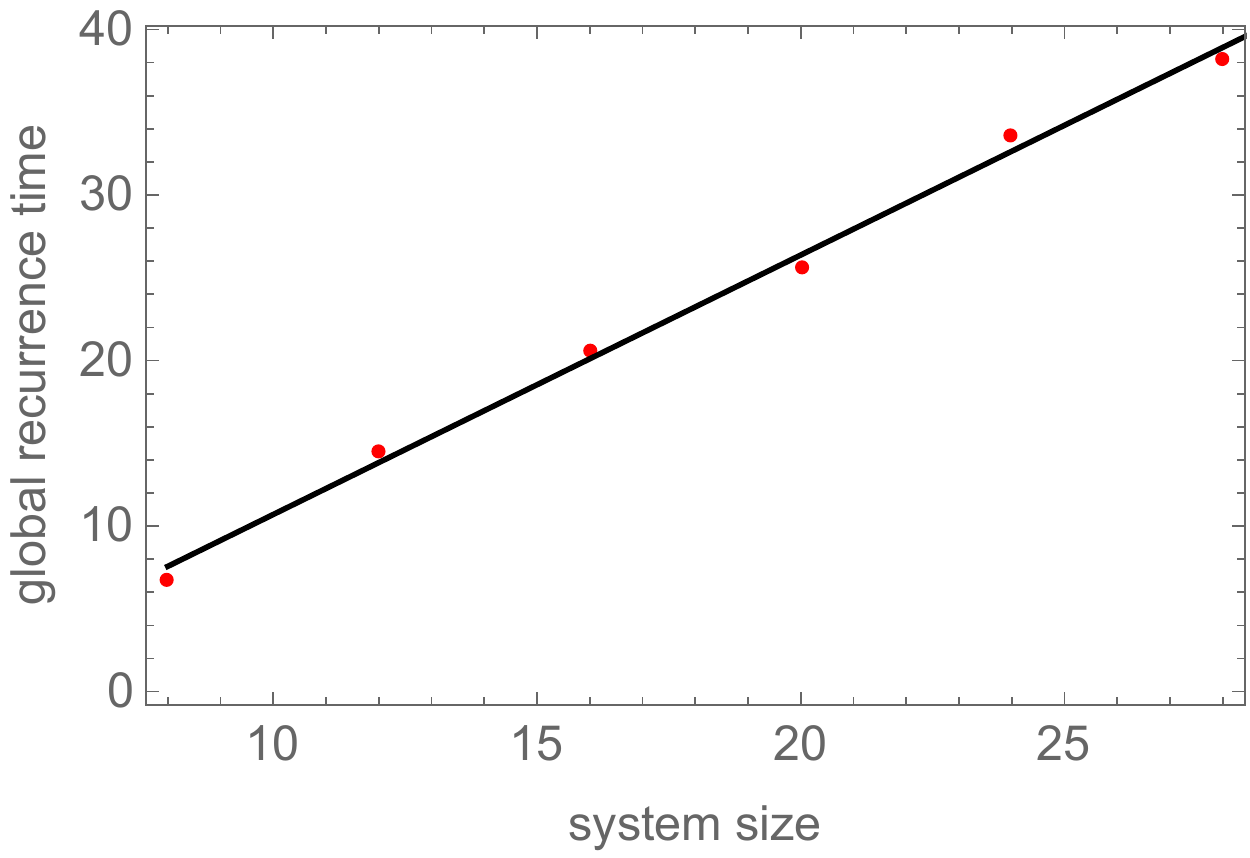}
\caption{Estimate of the logarithm of the global recurrence timescale vs. system size. The fit corresponds to the exponential quoted in the text.}
\label{Tg}
\end{center}
\end{figure}

\emph{OTOC for ``pair-hopping'' operator}.---Given the quasiparticle structure of this system, it seems natural that the $\sigma^x$ operator should scramble, as it usually changes the quasiparticle number. Correspondingly we have compared our results for this against the results for the OTOC $[Z_i(t), O_j(0)]$ with $O_j = \sigma^x_j \sigma^x_{j+1}$ and with $O_j = \sigma^-_j \sigma^-_{j+1} \sigma^+_{j+2} \sigma^+_{j+3}$. The results are shown in Fig.~\ref{otocvariants}; qualitatively these behave much like the results presented in the main text.

\begin{figure}[htbp]
\begin{center}
\begin{minipage}{0.25\textwidth}
\includegraphics[width = \textwidth]{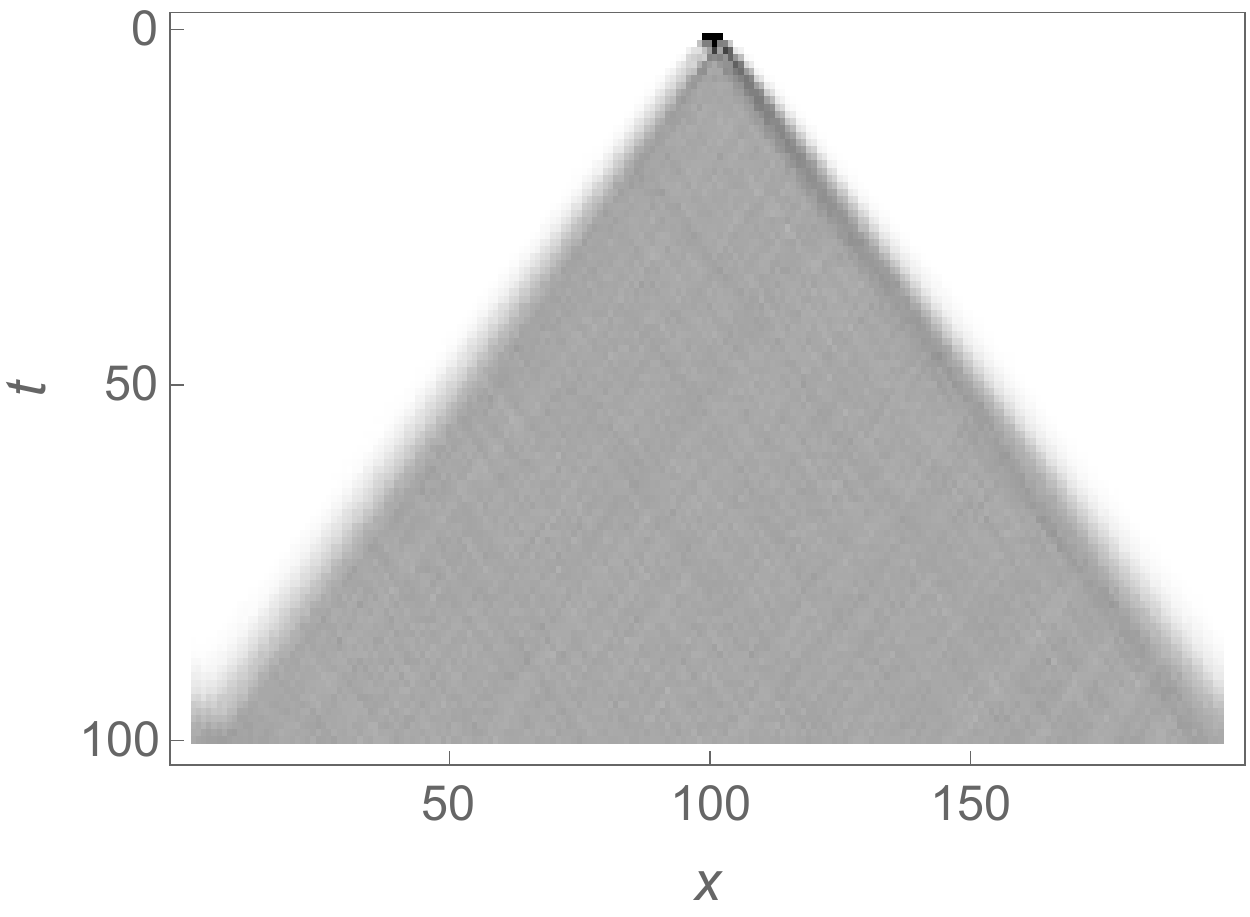}
\end{minipage}
\begin{minipage}{0.25\textwidth}
\includegraphics[width = \textwidth]{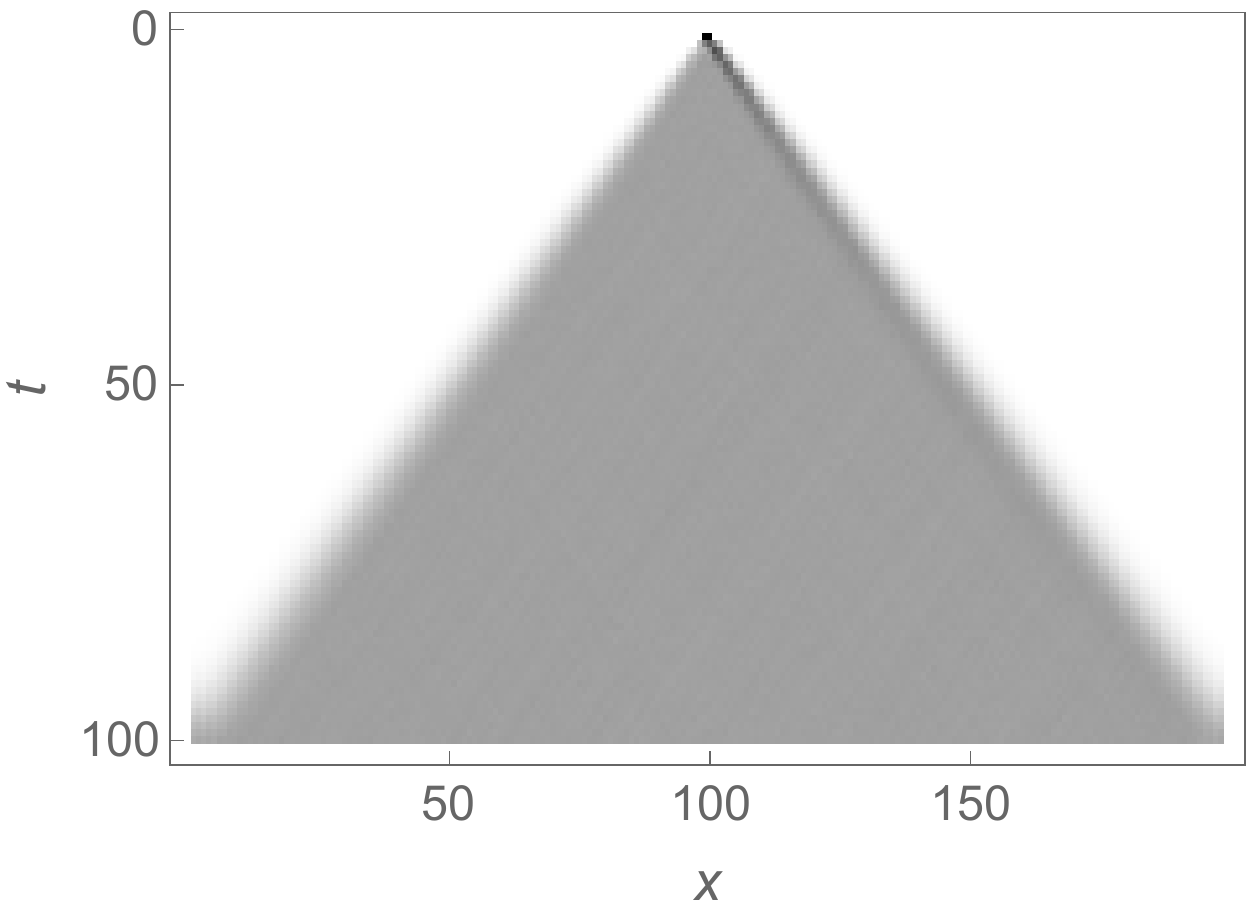}
\end{minipage}
\begin{minipage}{0.25\textwidth}
\includegraphics[width = \textwidth]{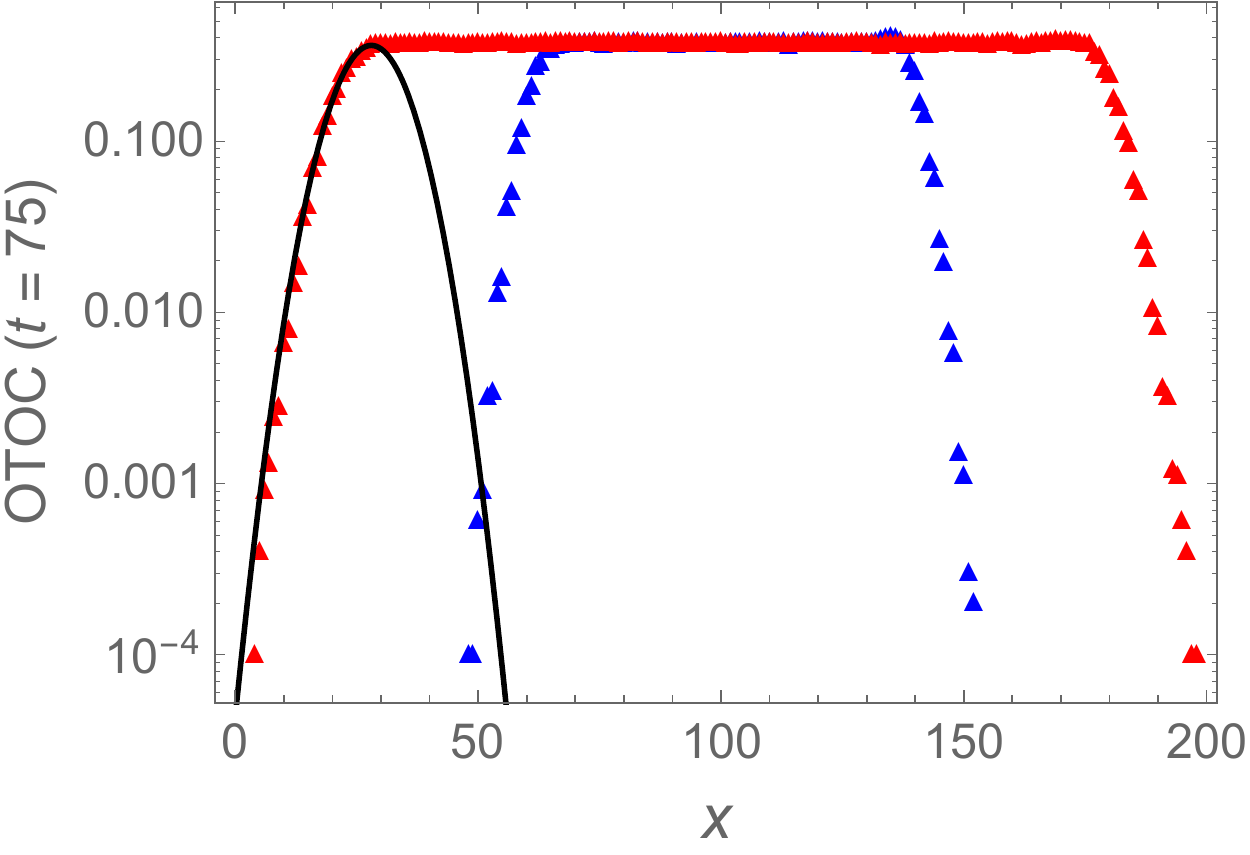}
\end{minipage}
\caption{Space-time growth of the fully averaged OTOCs $\langle [Z_i(t), O_j(0)]^2 \rangle$ with $O_j = \sigma^-_j \sigma^-_{j+1} \sigma^+_{j+2} \sigma^+_{j+3}$ (left) and $\sigma^x_j \sigma^x_{j+1}$ (center). Right: shape of the front for $\langle[Z_i(t), \sigma^x_j \sigma^x_{j+1}(0)]^2\rangle$, showing a clear Gaussian front.}
\label{otocvariants}
\end{center}
\end{figure}

\emph{OTOC in single product states and in eigenstates}.---We now present results for the OTOC in single product states and for the case where the quantum average is performed over a single random eigenstate. In the former case, the OTOC is always binary, but its pattern shows clear signs of the quasiparticle ``tracks'' going through the system (Fig.~\ref{specific_otocs}). 
When the OTOC is computed in a single eigenstate, as discussed in the main text, it spreads to some extent and then refocuses on a timescale of order $L$, as seen clearly in Fig.~\ref{specific_otocs}. 
While this single-eigenstate OTOC fills in the light-cone, the spread of the front is narrower than for the infinite-temperature-averaged OTOC, although our results are too noisy to draw any definite numerical conclusions about the shape of the front in this case. 

\begin{figure}[htbp]
\begin{center}
\begin{minipage}{0.3\textwidth}
\includegraphics[width =.8 \textwidth]{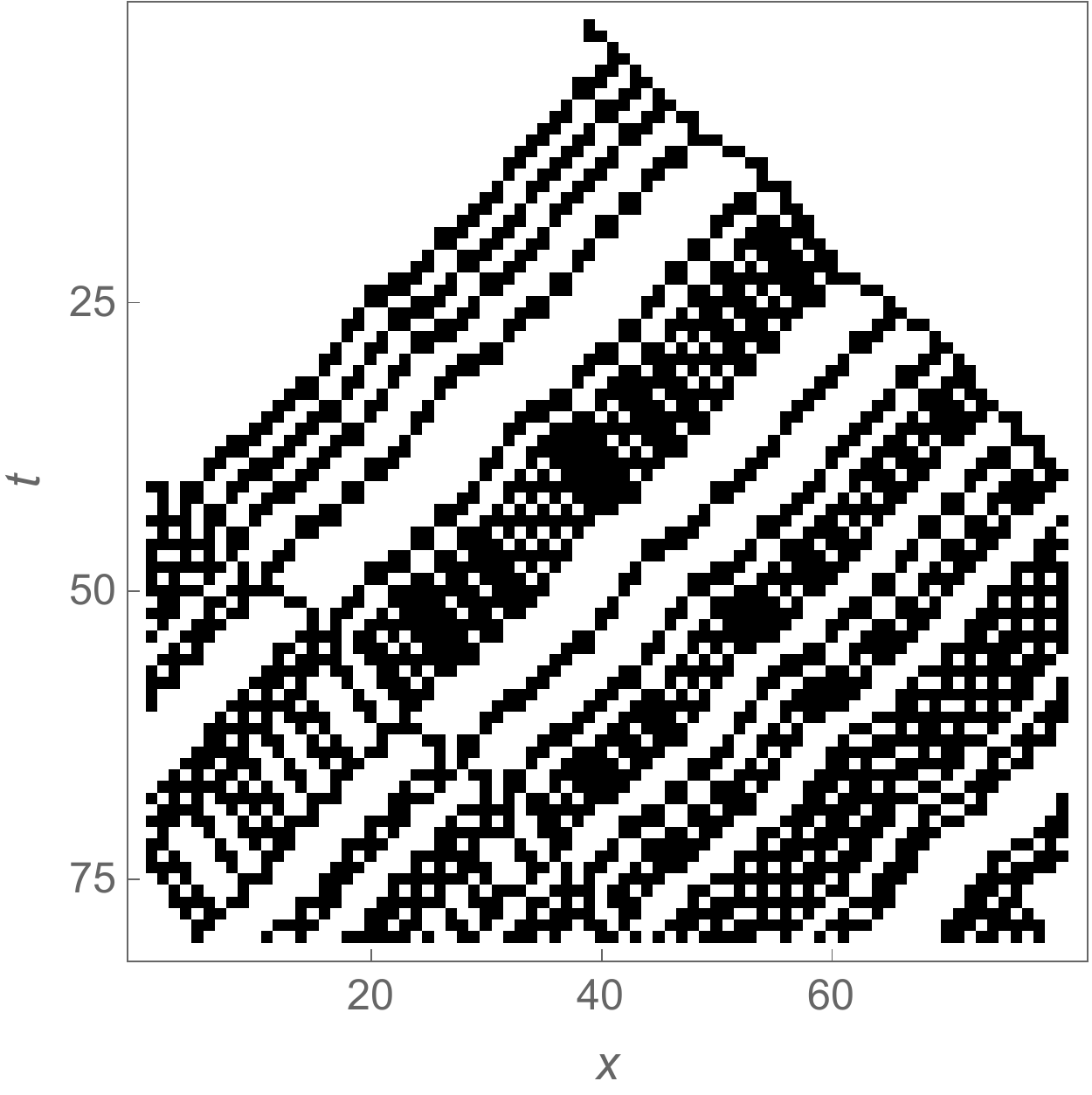}
\end{minipage}
%\begin{minipage}{0.3\textwidth}
%\includegraphics[width = \textwidth]{state_avg_otoc_40}
%\end{minipage}
\begin{minipage}{0.33\textwidth}
\includegraphics[width =\textwidth]{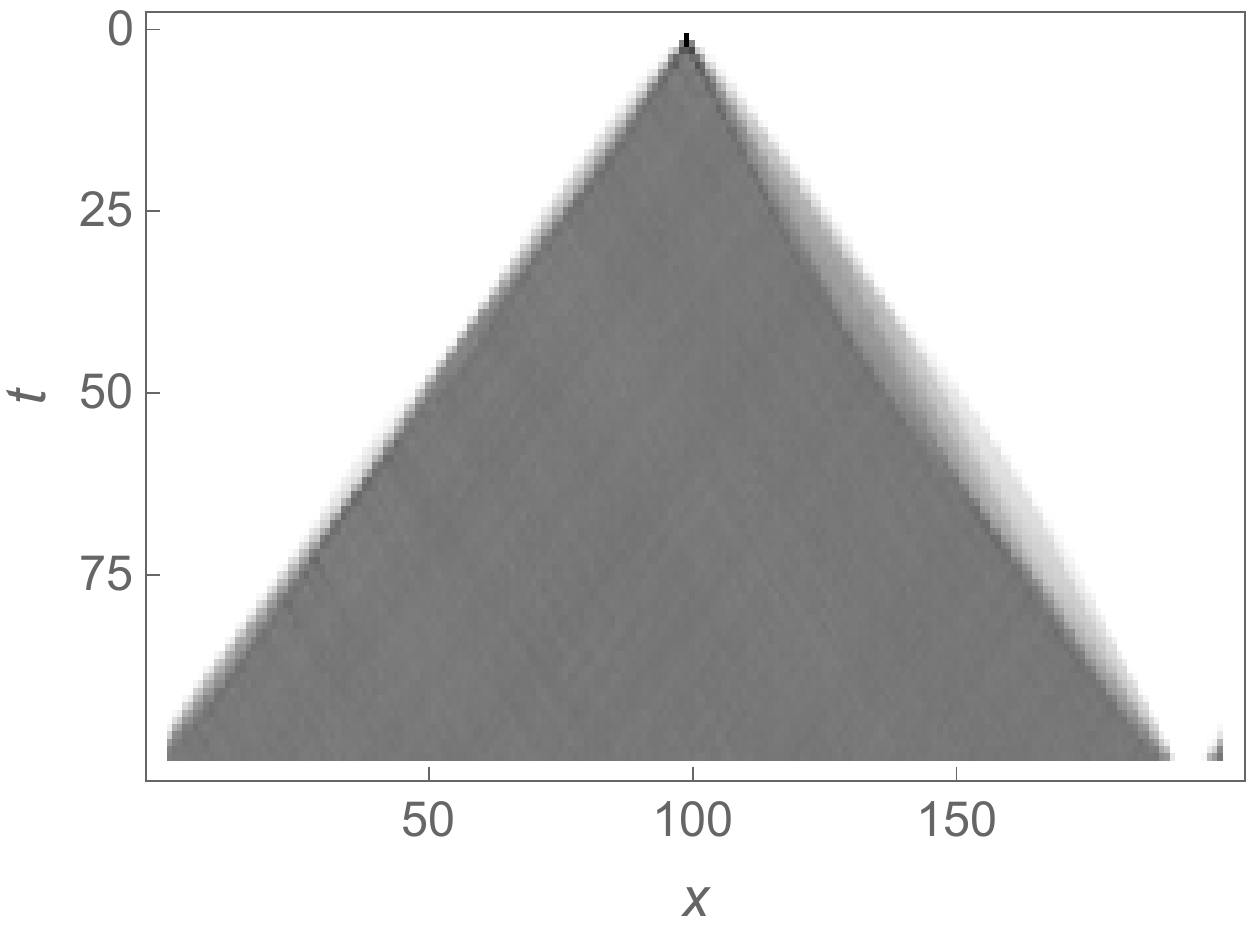}
\end{minipage}
\begin{minipage}{0.33\textwidth}
\includegraphics[width = \textwidth]{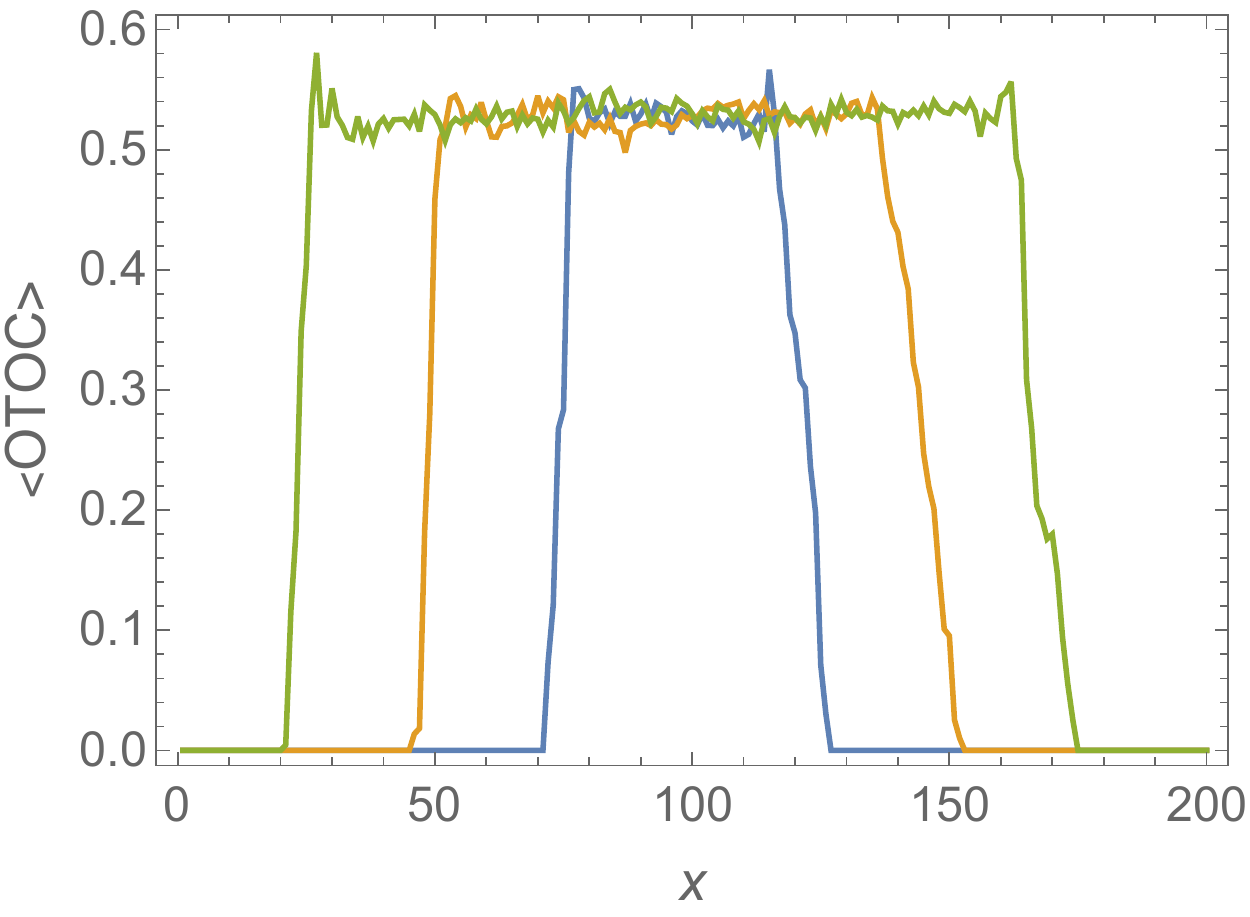}
\end{minipage}
\caption{Left: Space-time growth of OTOCs starting from a single product state for $L = 80$. Center: growth of OTOC starting from a single eigenstate for $L = 200$. Sections through this at $t = 25, 50, 75$ are shown in the right panel.}
\label{specific_otocs}
\end{center}
\end{figure}

\end{widetext}

\end{document}